\documentclass[12pt]{article}

\usepackage{amsmath,amsfonts,amssymb,appendix}
\usepackage{bm,booktabs,epstopdf}
\usepackage{color,colortbl,comment}
\usepackage{enumitem,enumerate,epsf,epsfig}
\usepackage{graphicx,graphics}
\usepackage{latexsym,lscape,longtable}
\usepackage{makeidx,mathrsfs,mathtools,multicol,multirow}
\usepackage[square,sort,comma,numbers]{natbib}
\usepackage{undertilde,url}
\usepackage{verbatim}
\usepackage{xcolor,xr-hyper}
\usepackage{hyperref}  

\newcommand{\0}{{\bf 0}}

\def\dfrac#1#2{{\displaystyle{#1\over#2}}}

\def\bse{\begin{eqnarray*}}
\def\ese{\end{eqnarray*}}
\def\be{\begin{eqnarray}}
\def\ee{\end{eqnarray}}
\def\bitm{\begin{itemize}}
\def\eitm{\end{itemize}}
\def\bq{\begin{equation}}
\def\eq{\end{equation}}
\def\bse{\begin{eqnarray*}}
\def\ese{\end{eqnarray*}}

\def\log{\hbox{log}}

\def\pr{\hbox{pr}}
\def\wh{\widehat}

\newcommand{\ben}{\begin{enumerate}}
\newcommand{\een}{\end{enumerate}}

\def\bSig\mathbf{\Sigma}

\newcommand{\uT} {\mbox{\boldmath$T$}}

\newcommand{\uV} {\mbox{\boldmath$V$}}

\newcommand{\uW} {\mbox{\boldmath$W$}}

\newcommand{\uX} {\mbox{\boldmath$X$}}
\newcommand{\ux} {\mbox{\boldmath$x$}}

\newcommand{\U} {\mbox{\boldmath$U$}}

\newcommand{\V} {\mbox{\boldmath$V$}}

\newcommand{\ubeta} {\mbox{\boldmath$\beta$}}
\newcommand{\ugamma} {\mbox{\boldmath$\gamma$}}

\newcommand{\uiota} {\mbox{\boldmath$\uiota$}}

\newcommand{\uDelta} {\mbox{\boldmath$\Delta$}}

\newcommand{\uSigma} {\mbox{\boldmath$\Sigma$}}

\newcommand{\btheta} {\mbox{\boldmath$\theta$}}
\newcommand{\bkappa} {\mbox{\boldmath$\kappa$}}

\newcommand{\tw} {\mbox{\text{w}}}
\newcommand{\e}{ { \mathbb{E}}}
\def\T{{ \mathrm{\scriptscriptstyle T} }}
\newcommand{\weakto}{\stackrel{\cal D}{\longrightarrow}}

\begin{document}

\title{Incorporating survival data into case-control studies with incident and prevalent cases}

\author{Soutrik Mandal$^1$, Jing Qin$^2$ \& Ruth M. Pfeiffer$^{1*}$}

\date{$^1$ National Cancer Institute, National Institutes of Health, Rockville, Maryland\\
$^2$ National Institute of Allergy and Infectious Diseases, National Institutes of Health, Bethesda, Maryland\\
\vspace{0.1in}
$^*$pfeiffer@mail.nih.gov}

\maketitle
 
\abstract{
Typically, case-control studies to estimate odds-ratios associating risk factors with disease incidence from logistic regression only include cases with newly diagnosed disease. Recently proposed methods allow incorporating information on prevalent cases, individuals who survived from disease diagnosis to sampling, into cross-sectionally sampled case-control studies under parametric assumptions for the survival time after diagnosis. Here we propose and study methods to additionally use prospectively observed survival times from prevalent and incident cases to adjust logistic models for the time between disease diagnosis and sampling, the backward time, for prevalent cases. This adjustment yields unbiased odds-ratio estimates from case-control studies that include prevalent cases. We propose a computationally simple two-step generalized method-of-moments estimation procedure. First, we estimate the survival distribution based on a semi-parametric Cox model using an expectation-maximization algorithm that yields fully efficient estimates and accommodates left truncation for the prevalent cases and right censoring.  Then, we use the estimated survival distribution in an extension of the logistic model to three groups (controls, incident and prevalent cases), to accommodate the survival bias in prevalent cases. In simulations, when the amount of censoring was modest, odds-ratios from the two-step procedure were equally efficient as those estimated by jointly optimizing the logistic and survival data likelihoods under parametric assumptions.  Even with 90\% censoring they were as efficient as estimates obtained using only cross-sectionally available information under parametric assumptions.  This indicates that utilizing prospective survival data from the cases lessens model dependency and improves precision of association estimates for case-control studies with prevalent cases.}


\section{Introduction}
 Case-control studies are   economical and therefore
  popular  for estimating the association of exposures with disease incidence for rare outcomes. 
They typically include only incident cases, i.e. individuals  with newly diagnosed disease   \cite{schlesselman82}.
 However, sometimes   subjects who developed the disease before the start of the study, termed ``prevalent cases'', are also available for sampling. Simply combining information from incident and prevalent cases leads to biased estimates of disease-exposure association, if the exposure of interest for disease incidence also impacts survival after disease diagnosis, as prevalent cases had to survive long enough to be available for sampling into the case-control study, i.e. their observations are left truncated. 
 
Maziarz et al  \cite{maziarz2019inference}  proposed a fully efficient method to estimate log-odds ratios of disease-exposure association when   prevalent cases are included in a cross-sectionally sampled case-control study.
They extended the exponential tilting (or density ratio) model of Qin \cite{qin1998inferences}
 for the disease-exposure relationship  to accommodate prevalent cases by correcting  for their survival bias  through a tilting term that depends on the distribution of the survival time following  disease onset. However, in  cross-sectionally sampled prevalent cases,  only the backward times, defined as the times from disease diagnosis to sampling, are observed and thus 
 a fully parametric model for  the  distribution of the survival time following disease onset is required     for identifiability.
 
Often   prospective follow-up information on the time from disease diagnosis to death for incident and prevalent cases is available. In this paper,  we  propose and study methods to utilize this  prospective information  to relax  parametric assumptions   
and     to estimate  the tilting term in the logistic model for prevalent cases based on  the   popular semi-parametric  Cox proportional hazards model 
  \cite{cox72}.
 Information on survival after disease onset  can typically be obtained   readily  when cases and controls are sampled from a well-defined cohort, and  investigators periodically update the vital status of cohort members by linking to national databases such as  the US National Death Index. This is the setting of the study that motivated our work,  
 a case-control study that included controls, and incident and  prevalent breast cancer cases sampled from within the US Radiologic Technologists Study (USRTS) cohort to estimate the association between selected single nucleotide polymorphisms (SNPs) and  breast cancer risk. 
A second  important setting where prospective survival information    is  commonly available is  for  case-control studies that use data extracted from health insurance claims databases and electronic medical  records. 
A related example is the widely used  Surveillance, Epidemiology and End Results (SEER)-Medicare database,  created by linking   SEER cancer registries with Medicare claims from the time of a person's Medicare eligibility until death, that also contains   
 a 5\% random sample of Medicare beneficiaries from which controls  can be obtained 
 (https://healthcaredelivery.cancer.gov/seermedicare/).

In this paper, we derive a profile likelihood that combines the retrospectively sampled case-control data    and the prospective survival data for cases,   while accommodating left truncation for the prevalent cases and right censoring. When the survival model is parametrically specified, we jointly maximize this likelihood for the 
logistic parameters and the parameters in the survival model. 
However, for  a semi-parametrically specified survival distribution the joint estimation is computationally prohibitively complex. We thus develop a two-step generalized methods of moments procedure to estimate log-odds ratios for association of  exposures with disease incidence from  case-control studies that also include prevalent cases when prospective survival data on cases are available.
  In the first step of the  procedure  we estimate  the survival distribution based on a  Cox proportional hazards model using   prospective follow-up time information available on the cases.  
We propose   an expectation-maximization (EM) algorithm that builds on work by Liu et al   \cite{liu2016semiparametric} and Qin et al  \cite{qin2011maximum}, 
to obtain maximum likelihood estimates (MLEs) of the  Cox model parameters for right-censored, left-truncated  data  that are fully  efficient.  We also compare these results with those from   maximizing the partial likelihood accommodating left truncation of the prevalent cases  and right censoring in the data, that can be implemented using standard statistical software.  
In the second step, we  use the estimated survival distribution  in  
the  semi-parametric profile likelihood for controls,  incident and prevalent cases and estimate log-odds ratios for disease-exposure associations. Section \ref{sec:twostep} details the two-step procedure after introducing the model and notation in Section  \ref{sec:logsec}.

 We derive the asymptotic properties and study the performance of the proposed method using simulations with varying amounts of censoring of the prospective survival data (Section \ref{sec:sim}). We  compare the efficiency of log-odds ratio estimates that utilize prospective survival information with the  efficiency  of estimates from Maziarz et al \cite{maziarz2019inference}. This may help decide whether collecting   prospective information on the cases in addition to cross-sectional data on covariates and the backward time for prevalent cases increases the precision of the resulting estimates sufficiently to warrant the cost for any additional data collection.   
We  illustrate the methods by analyzing the USRTS breast cancer case-control study, that included  incident and prevalent cases  (Section \ref{sec:data}).
 Section \ref{sec:disc} closes with a discussion.


\section{Models and likelihood}%
\label{sec:logsec}

First  we summarize the data and models used  in Maziarz et al \cite{maziarz2019inference} and then incorporate prospective information on the time from disease diagnosis to death for the cases sampled into the case-control study. 
 
\subsection{Background: Semi-parametric model for case-control studies with incident and prevalent cases}\label{sec:mazback}

  Let $D$ denote the disease indicator, with $D=1$  for those with disease (cases) and $D=0$ for those without (controls), and let $\uX$ denote a vector of covariates. 
\subsubsection{Exponential tilting models}\label{exptiltsec}

We assume   in the population  $\uX$ is associated with incident disease through  the  logistic model  
\be\label{eq:logistic}
P(D=1|\uX=\ux)= \frac{\exp(\alpha_0+\ux^T\ubeta)}{1+\exp(\alpha_0+\ux^T\ubeta)},
\ee
where $\alpha_0$ is the intercept term and $\ubeta$ is the vector of log-odds ratios.

The marginal probability of disease in the population is  $\pi=P(D=1)=\int P(D=1|\ux)f(\ux)d\ux$ where $f(\ux)=dF(\ux)/d\ux$ is the unspecified density corresponding to  the cumulative distribution function  $F(\ux)$  of $\uX$.
In retrospectively sampled case-control studies   we observe the  conditional densities $f_0(\ux) = f(\ux|D = 0)$  for  controls, and 
$f_1(\ux) = f(\ux|D = 1)$ for   incident cases. 
Under the population model (\ref{eq:logistic}), these two densities are related through the   exponential tilting (or density ratio) model  \cite{qin1998inferences},
\be\label{eq:exptilt}
f_1(\ux)=\frac{\exp(\alpha_0+\ux^T\ubeta)}{1+\exp(\alpha_0+\ux^T\ubeta)}\frac{f(\ux)}{\pi}= f_0(\ux)\exp(\alpha^*+\ux^T\ubeta)
\ee
where $\alpha^*= \alpha_0+\log\{(1-\pi)/\pi\}$.  
 
Maziarz et al \cite{maziarz2019inference} extended  model (\ref{eq:exptilt}) to   accommodate covariates $\uX$ from prevalent cases. 
However, only those prevalent cases are observed  
whose  backward time $A$, i.e. the  time between disease diagnosis and sampling into the case-control study,  is shorter than  the time $T$ from  diagnosis to death, i.e. they are subjected to left truncation.  
 Thus the joint distribution of the  observed data $(\uX, A|A<T,D=1)$ for prevalent cases    is 
\bse\label{eq:jtXA}
f(\uX=\ux,A=a|D=1,T>A)=f(\uX=\ux|D=1,T>A)f(A=a|\uX=\ux,D=1,T>A).
\ese
If the disease incidence is stationary over time, the backward time $A$ has a uniform distribution in some interval $[0, \xi]$ (see detailed explanation in Maziarz et al \cite{maziarz2019inference} and our comment on this assumption in the Discussion). 
Further assuming that the time to disease onset and $T$  are independent,  
 the density of $\uX$ for prevalent cases is 
\bq\label{eq:f2}
f_2(\ux)=f(\ux|D=1,T>A)=f_0(\ux)\exp\{\nu^*+\ux^T\ubeta+\log\mu(\ux,\bkappa)\}
\eq
where $  S(t|\ux,\bkappa) = P(T>t|\ux,\bkappa)$ denotes the survival distribution of $T$  with  parameters $\bkappa$, 
\bq
\label{eq:mu0}\mu(\ux,\bkappa)=\int_0^\xi S(a|\ux,\bkappa)da,
\eq
 $\nu^*=\alpha^*-\log\{\int_{\mathcal X}\mu(\ux,\bkappa)f_1(\ux)d\ux\}$  and $\mathcal X$ is the support of $\uX$.
As $A$ is independent of $\uX$ and $T$, the conditional density of $A$, for $a \in [0,\xi]$, is
\be\label{eq:AgivenX}
f_A(A=a|\uX=\ux,D=1,T>A)=\frac{f(A=a)P(T>a|\uX=\ux,D=1)}{P(T>A|\uX=\ux,D=1)}=\frac{S(a|\ux,\bkappa)}{\mu(\ux,\bkappa)},
\ee
and $0$ for $a \notin [0,\xi]$.
If only the backward times $A$  are observed,  but not the actual survival times, $S$  needs to be specified fully parametrically to ensure identifiability.

\subsubsection{Profile log-likelihood}
\label{sec:logl}

 Using the exponential tilting models (\ref{eq:exptilt}) and (\ref{eq:f2}), and the backward time distribution (\ref{eq:AgivenX}), the likelihood for the cross-sectionally observed data for the  controls and the two case groups is
\bse\label{eq:lik}
\mathcal L = \prod_{i=1}^N f_0(\ux_i)\prod_{i=n_0+1}^{n_0+n_1}\exp(\alpha^*+\ux^T_i\ubeta)\prod_{i=n_0+n_1+1}^N\exp\{\nu^*+\ux^T_i\ubeta+\log\mu(\ux_i,\bkappa)\}\frac{S(a_i|\ux_i,\bkappa)}{\mu(\ux_i,\bkappa)},
\ese
where $(\ux_1, \dots, \ux_{n_0})^T$ are the covariates for the $n_0$ controls, $(\ux_{n_0+1}, \dots, \ux_{n_0+n_1})^T$ the covariates for the $n_1$ incident cases, and $(\ux_{n_0+n_1+1}, \dots, \ux_N)^T$ and $(a_{n_0+n_1+1}, \dots, a_N)^T$ the covariates and backward times for the $n_2$ prevalent cases, with $N = n_0 + n_1 + n_2$.
To ensure that $f_i, i = 0, 1, 2$ are distributions, 
the  $p_i = f_0(\ux_i) = P(\uX = \ux_i)$, $i = 1, \dots,N$, are estimated empirically  after imposing the following constraints:
$\sum_{i=1}^N p_i=1$, $p_i\geq 0$; $\sum_{i=1}^N p_i\exp(\alpha^*+\ux^T_i\ubeta)=1$; and $\sum_{i=1}^N p_i\exp\{\nu^*+\ux^T_i\ubeta+\log\mu(\ux_i,\bkappa)\}=1$ via Lagrange multipliers. After maximizing the log-likelihood for $p_i$ subject to the constraints, the profile log-likelihood for the remaining parameters $(\alpha,\nu,\ubeta,\bkappa)$ is  
\be
\label{eq:loglik}
l_p(\alpha,\nu,\ubeta,\bkappa)&=& -\sum_{i=1}^N\log[1+\exp(\alpha+\ux^T_i\ubeta)+\exp\{\nu+\ux^T_i\ubeta+\log\mu(\ux_i,\bkappa)\}]+\sum_{i=n_0+1}^{n_0+n_1}(\alpha+\ux^T_i\ubeta)\nonumber\\
&& +\sum_{i=n_0+n_1+1}^N\Big[\nu+\ux^T_i\ubeta+\log\mu(\ux_i,\bkappa)+\log\Big\{\frac{S(a_i|\ux_i,\bkappa)}{\mu(\ux_i,\bkappa)}\Big\}\Big],
\ee
where $\alpha=\alpha^*+\log(n_1/n_0)$ and $\nu=\nu^*+\log(n_2/n_0)$.

\subsection{Incorporating  prospective follow-up information on the cases}  
\label{sec:pros}

We now assume that in addition to case-control status, covariates and the backward time for prevalent cases,  prospective follow-up information on the time from disease diagnosis to death is observed on all cases in  the case-control study. This  is the setting of our motivating study that sampled incident and prevalent  breast cancer cases   and controls from  the  USRTS cohort to assess the association of breast cancer risk with SNPs in select genes. Investigators regularly  link USRTS  cohort members with the US national death index (NDI), to update  vital status data.

\begin{figure}[ht]
\hspace*{-3cm}
\includegraphics[scale=0.66]{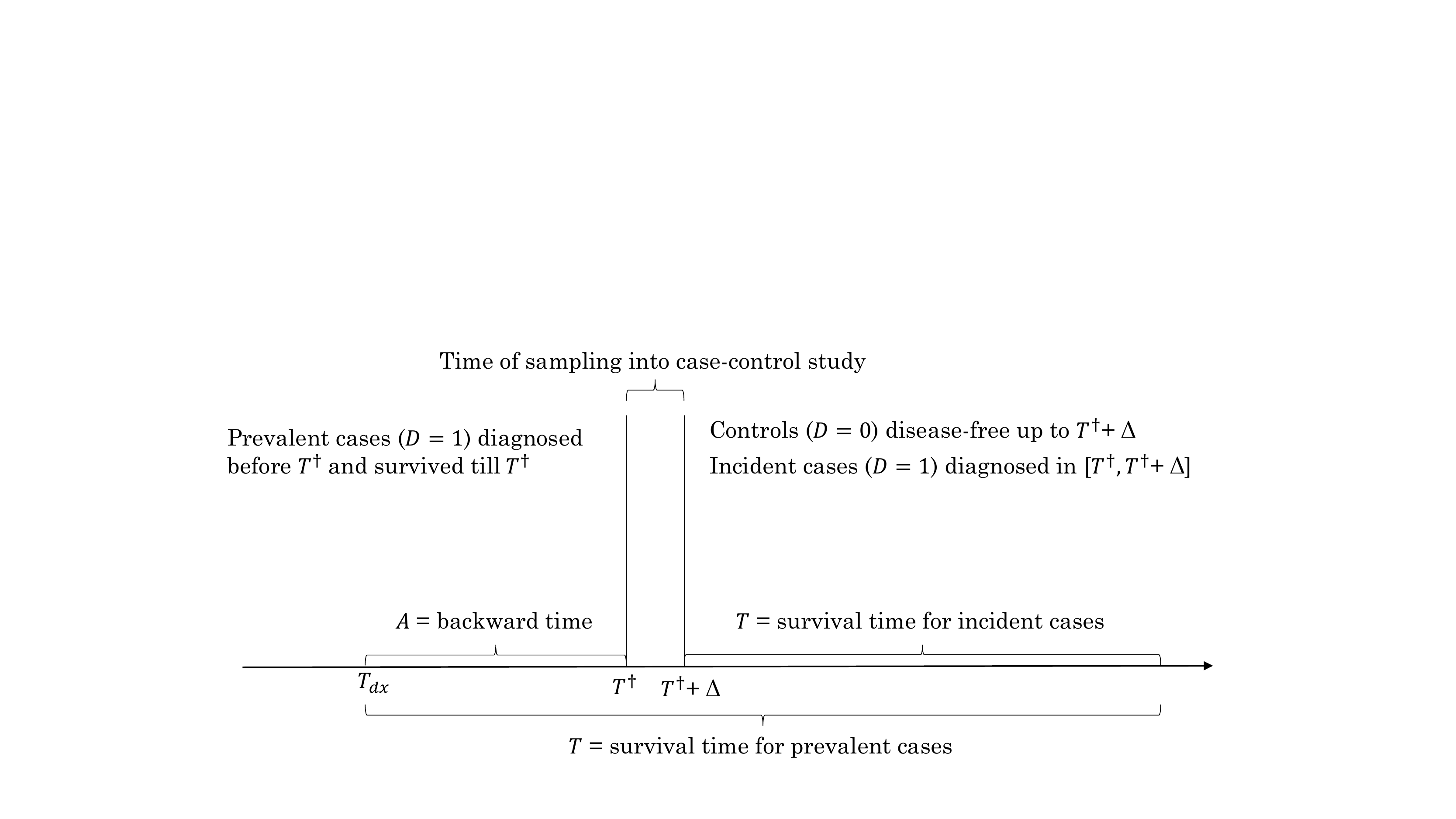}
\caption{Sampling and survival times for controls, incident and prevalent cases in case-control study. $[T^\dagger, T^\dagger+\Delta]$ denotes the case-control sampling period, $T_{dx}$ is the diagnosis time for prevalent cases and $T$ the time from disease diagnosis to death. 
}\label{plot1}
\end{figure}

Recall that  $T$ denotes the time from  disease diagnosis to death and let $C$ denote  the censoring time.
We define the observed time to be $Y= \min(T, C)$ and  the event indicator  $\delta=1$, if $Y=T$ and  $\delta=0$ otherwise.  The data observed for an incident case are $O= (Y, \delta, \uX=\ux)$ and for 
 a prevalent case   $O= (Y,   \delta, \uX=\ux|A<T)$, i.e. the survival times are   left-truncated in addition to being right-censored.  We assume that $(T,A)$ and $C$ are conditionally independent given $\uX$, and, as before, $T$ and $A$ are independent given $\uX$.
Figure \ref{plot1} summarizes  the sampling scheme for all individuals in the study and the available prospective survival information.

 
We  model    the dependence of $S$ on $\uX$  using a  Cox proportional hazards model  \cite{cox72},
 and thus $S(t|\uX,\bkappa)= \exp\{ - \Lambda(t|\uX,\bkappa)\}$     with
 \begin{equation}\label{eq:prop} 
d\Lambda(t|\uX=\ux,\bkappa)= \lambda(t|\uX=\ux,\bkappa)=\lambda_0(t) \exp(\ux^T\ugamma),  \end{equation}
  where $\lambda_0$ is the   baseline hazard function that depends only on time, and $\bkappa=(\lambda_0, \ugamma)$.

Letting the indicator variable $R= 1$ for a prevalent case, and $R= 0$ for an incident case, and 
$g(t|\uX,\bkappa) = -d S(t|\uX,\bkappa)/dt$ denote the density corresponding to $S$,    the likelihood for the survival   data for the $n_1$ incident and $n_2$ prevalent cases is
 \be \label{eq:survivalL}
 L_S(\bkappa) = \prod_{i=n_0+1}^{N}   g^{\delta_i}(Y_i|\ux_i, \bkappa)S^{(1-\delta_i)}(Y_i|\ux_i, \bkappa) 
 \left\{\mu(\ux_i, \bkappa)\right\}^{-R_i}.
\ee
 Combining the  log-likelihood corresponding to (\ref{eq:survivalL}) for the prospective survival data with the profile log-likelihood $l_p(\alpha,\nu,\ubeta,\bkappa)$ in  (\ref{eq:loglik}) for the case-control data and the backward time on the prevalent cases yields the full data profile log-likelihood 
 \be
\label{eq:fullL}
l(\alpha, \nu, \ubeta, \bkappa)&=& -\sum_{i=1}^N\log[1+\exp(\alpha+\ux^T_i\ubeta)+\exp\{\nu+\ux^T_i\ubeta+\log\mu(\ux_i,\bkappa)\}]\nonumber \\  
&& +\sum_{i=n_0+1}^{n_0+n_1}(\alpha+\ux^T_i\ubeta)+
\sum_{i=n_0+n_1+1}^N\Big[\nu+\ux^T_i\ubeta+ \log\Big\{\frac{S(a_i|\ux_i,\bkappa)}{\mu(\ux_i,\bkappa)}\Big\}\Big]\nonumber \\  
&& +\sum^{N}_{i=n_0+1}[\delta_i \log g(y_i|\ux_i,\bkappa) +(1-\delta_i)\log S(y_i|\ux_i,\bkappa)].
\ee
 Under a parametric model for $\lambda_0(t)$ in (\ref{eq:prop}) 
 the above profile log-likelihood can be maximized jointly for all parameters using standard optimization. 
E.g.  when  $\lambda_0 $ is a Weibull hazard with  shape and scale parameters $\kappa_1$ and $\kappa_2$,  respectively,   $S(t|\ux,\bkappa)= \exp\{-(t/\kappa_2)^{\kappa_1} \exp(\ux^T\ugamma)\}$ and  $\mu(\ux,\bkappa)= \Gamma(\kappa^{-1}_1)/(\kappa_1\kappa_3^{1/\kappa_1})\Big\{\Gamma^{-1}(\kappa^{-1}_1)\int^{\kappa_3\xi^{\kappa_1}}_0 \exp(-u)u^{(1/\kappa_1-1)}du\Big\}$ with $\kappa_3= \kappa_2^{-\kappa_1}\exp(\ux^T \ugamma)$.

However, for  a general unspecified baseline hazard function  $\lambda_0(t)$  in  (\ref{eq:prop})  that one wishes to estimate non-parametrically, optimizing (\ref{eq:fullL})
 becomes computationally extremely   difficult, even though it is theoretically identifiable. Instead, we propose a two-step generalized method of moment approach for estimation that we discuss next.

\section{Two-step parameter estimation} 
\label{sec:twostep}

We now propose and study a two-step generalized method of moments approach to estimating $(\alpha, \nu, \ubeta, \bkappa)$ 
when  $S$ is modeled based on the Cox proportional hazard model, i.e. 
$ \lambda(t|\ux, \bkappa)=\lambda_0(t) \exp(\ux^T\ugamma)$, where $\lambda_0$ is an unspecified baseline hazard function  \cite{cox72}.   First, we  estimate  $\bkappa=(\lambda_0,\ugamma)$  semi-parametrically using  the prospective survival information from   incident and/or prevalent cases. 
Then we plug  $\wh{\bkappa}$
into (\ref{eq:lik})  and maximize the pseudo-log-likelihood as a function of the remaining parameters $(\alpha, \nu, \ubeta)$.

\subsection{{\it Step 1.} Estimate $\bkappa=(\lambda_0,\ugamma)$   from prospective  follow-up data for
incident and prevalent   cases}\label{sec:sandmu}
 
We adapt  an  EM algorithm proposed by  Qin et al
 \cite{qin2011maximum}  and further modified by Liu et al  \cite{liu2016semiparametric}    to estimate $\lambda(t|\ux,\bkappa)$ in (\ref{eq:prop}) when  follow-up information from prevalent and incident  cases is available, to obtain  fully efficient estimates of $\bkappa$. For comparison  we also estimate $\bkappa$ based on  the standard Cox partial likelihood with left truncation.

 {\it a) Estimating $\bkappa$ via an EM algorithm} 

The basic idea for the EM algorithm is  that for the $i$th  prevalent case that is observed,  $m_i$ cases were left-truncated, i.e. unobserved. The ``missing  data'' for the $i$th prevalent case are  thus  $O^*_i= \{(T^*_{i1},A^*_{i1}),\dots, (T^*_{im_i},A^*_{im_i})\}$ where $T^*_{il}$ and $A^*_{il}$ are the survival and backward times, respectively, for the $l$th unobserved prevalent case with  $T_{i1}^*<A_{i1}^*$. 
 The complete   data for each prevalent case are  then  $(O, O^*)$.

 Let  $\lambda_j= \lambda_0(t_j)$, where $t_j$, $j= 1,\dots,k$ are the observed  event times  for all cases. The  complete data log-likelihood for incident and prevalent cases, based on (\ref{eq:survivalL}) and (\ref{eq:prop})     is  
\bse
l_c(\bkappa)= l_c(\ugamma,\lambda_0)&=&\sum_{j=1}^k \sum_{i=1}^{n_1+n_2} \Big[ I(Y_i=t_j)\{\delta_i(\log\lambda_j+\ux^T_i \ugamma)-\exp(\ux^T_i\ugamma)\sum_{p=1}^j\lambda_p\}\\
&& + \sum_{l=1}^{m_i} r_i I(T^*_{il}=t_j)\{\log\lambda_j+\ux^T_i\ugamma-\exp(\ux^T_i\ugamma)\sum_{p=1}^j\lambda_p\}
\Big].
\ese
$I$ denotes the indicator function that is $1$ if the argument is true and $0$ otherwise.
Conditional on the observed data  $O_i$  for the $i$th subject, we write the expectation in the E-step as
\bse
\tw_{ij}= E\Big[\sum_{l=1}^{m_i}I(T^*_{il}=t_j)|O_i\Big]= \frac{\wh\xi}{\upsilon_i}\Big(1-\frac{t_j}{\wh\xi}\Big)\omega_{ij}
\ese
where $\omega_{ij}= \lambda_j \exp(\ux^T_i\ugamma)\exp\{-\sum_{l=1}^j\lambda_l \exp(\ux^T_i\ugamma)\}$, $\upsilon_i= \sum_{j=1}^k t_j \omega_{ij}$,  
  and following Qin et al \cite{qin2011maximum}, $\wh{\xi}= t_k$.   
In the M-step, we maximize the expected complete-data log-likelihood function conditional on the observed data, 
\be\label{eq:Q}
Q(\bkappa|\bkappa^{(u)})&=&\sum_{j=1}^k \sum_{i=1}^{n_1+n_2} \Big[I(Y_i=t_j)\{\delta_i(\log\lambda_j+\ux^T_i\ugamma)-\exp(\ux^T_i\ugamma)\sum_{p=1}^j\lambda_p\}\nonumber\\
&& +\tw_{ij}^{(u)}r_i\{\log\lambda_j+\ux^T_i\ugamma-\exp(\ux^T_i\ugamma)\sum_{p=1}^j\lambda_p\}\Big].
\ee
 
 Estimates $\wh{\ugamma^{(u)}}$ can be computed by fitting a weighted Cox regression model, e.g.  using the {\it coxph} function  in {\it R},
{\it coxph $($Surv $(\uT_{(n_1+n_2)k}, \uDelta_{(n_1+n_2)k})\sim \uX_{(n_1+n_2)k}$, weights= $\uW_{(n_1+n_2)k})$}, where  $\uT_{(n_1+n_2)k}= (t_1,\dots,t_k,\dots,t_1,\dots,t_k)$, $\uX_{(n_1+n_2)k}= (\ux_{n_0+1},\dots,\ux_{n_0+1},\dots,\ux_N,\dots,\ux_N)$, $\uDelta_{(n_1+n_2)k}= (1,\dots, 1)$, and $\uW_{(n_1+n_2)k}= (\tw^{(u)}_{11},\dots,\tw^{(u)}_{1k},\dots,\tw^{(u)}_{(n_1+n_2)1},\dots,\tw^{(u)}_{(n_1+n_2)k})$  \cite{liu2016semiparametric,qin2011maximum}.
Estimates  $\wh \lambda_j(\ugamma)$ in (\ref{eq:Q}) have closed-form solutions, 
\bse
\wh\lambda^{(u)}_j(\ugamma)= \frac{\sum_{i=n_0+1}^N\{\tw^{(u)}_{ij}r_i+I(Y_i=t_j)\delta_i\}}{\sum_{i=n_0+1}^N\sum_{l=j}^k\{\tw^{(u)}_{il} r_i+
I(Y_i=t_l)\}\exp(\ux^T_i\ugamma^{(u)})}.
\ese
 
{\it Remark:} Our derivations above assumes that the backward time $A$ has a uniform distribution, however, the algorithm can be extended to other parametric distributions for $A$, as shown in Supplemental Material,
   if stationarity of disease incidence in the underlying  population is an unreasonable assumption. 

\noindent   {\it b) Cox partial likelihood with left truncation} 
  
While estimates $\wh{\bkappa}$ from the EM algorithm are more efficient   \cite{liu2016semiparametric}, one could also  estimate   $\bkappa$ using    the standard Cox   partial likelihood  \cite{cox72, cox75}.
For  individual $i$ the   counting process   $N_i(t)$  is defined as $N_i(t)=I(Y_i \leq t, \delta_i=1), t \geq 0.$
Left truncation of the prevalent cases is accommodated in   the   ``at risk process"    $Z_i(t)=I(A_i <  t \leq  Y_i)$, where $A_i$ is the backward time if $i$ is  a prevalent case  and  $A_i=0$ for an  incident case.
$Z(t)$ is  not monotone decreasing with $t$  as prevalent cases are at risk only since   time $A$.  
Letting $dN_i(t) = N_i(t) - N_i(t-)$ denote the increment of $N_i$ at  time  $t$, 
 the   score functions based on the partial likelihood for $\ugamma$ are
\begin{align}
\boldmath{U}_{11}(\ugamma) = \sum_{i=1}^n   \sum_{t \geq 0}\bigg\{ \ux_i - \dfrac{\wh{S}^{(1)}(t;\ugamma)}{\wh{S}^{(0)}(t; \ugamma)}\bigg\}\mathrm{d}N_i(t)=\0,  
\label{eq:u11}
\end{align}
with $\wh{S}^{(r)}(t;\ugamma)=\sum_{j=1}^n Z_{j}(t) \exp(\ugamma^\text{T} \ux_j)\ux_j^{\otimes r}$, 
where for a column vector $a$, $a^{\otimes 0}=1, a^{\otimes 1}=a$, $a^{\otimes 2}=aa^{T}$ and $n= n_1+n_2$.    
 Given $\widehat{\ugamma}$, the estimating equations for $\lambda_0(t)$ are
\begin{align}
\boldmath{U}_{12}\{\lambda_0(t)\mid{\widehat{\ugamma}}\}= \sum_{i=1}^n 
\big\{\mathrm{d}N_i(t)-\lambda_0(t) Z_{i}(t)\exp(\widehat{\ugamma}^\text{T} \ux_i)\big\}=\0,  
\label{eq:u12}
\end{align}   
resulting in  the Breslow estimate of the cumulative  baseline  hazard at time $t$,  
\begin{equation}
\wh{\Lambda}_{0}(t) =  \int_0^t  \wh{\lambda}_{0}(s)ds =
\sum_{t_k \leq t}  \frac{\sum_{i=1}^n   dN_{i}(t_k)}{\sum_{i=1}^n  Z_i(t_k) \exp ( \wh{\ugamma}' \ux_i)},
\label{eq:surv_base}
\end{equation}
        with $0= t_0 <\dots<t_k$ denoting the   observed event times for  all cases  \cite{breslow72,aalen78}.  
 
\subsection{Estimating  $\mu$}
\label{sec:mu}
 
 For a given covariate $\uX= \ux$, we    use    $\wh{\bkappa}$ and  $\wh{\xi}= t_k$ in expression 
(\ref{eq:mu0}),  and obtain  
\be\label{eq:mu}
\mu(\ux,\wh{\bkappa}) &=& 
\int_0^{t_k\wedge \wh{\xi}} \exp\{-\wh{\Lambda}_0(t) \exp(\ux^T\wh{\ugamma})\}dt  \nonumber\\ 
&=& \sum_{j=1}^kI(t_{j-1}\leq\wh{\xi}) \{ (t_j\wedge\wh\xi) -t_{j-1} \} \exp\{-\wh{\Lambda}_0(t_{j-1})\exp(\ux^T\wh{\ugamma})\}.
\ee

\subsection{{\it Step 2.}  Estimate    $\btheta=(\alpha, \nu, \ubeta)$ given $\wh{\bkappa}$.}
 
We now treat   $\mu(\uX,\wh{\bkappa})$ in (\ref{eq:mu})   as a known function of $\uX$    and estimate the  remaining parameters $(\alpha,\nu,\ubeta)$ by maximizing the  pseudo log-likelihood 
\be
\label{eq:twostepL}
l(\alpha, \nu, \ubeta| \wh{\bkappa})&=& -\sum_{i=1}^N\log[1+\exp(\alpha+\ux^T_i\ubeta)+\exp\{\nu+\ux^T_i\ubeta+\log  \mu(\ux_i,\wh{\bkappa})\}]\\ \nonumber
&& +\sum_{i=n_0+1}^{n_0+n_1}(\alpha+\ux^T_i\ubeta)+
\sum_{i=n_0+n_1+1}^N (\nu+\ux^T_i\ubeta).   
\ee

{\it Theorem:} Denote the   estimator that maximizes   (\ref{eq:twostepL}) by $\wh{\btheta}=(\alpha, \nu, \beta)^T$ and the true value by $\btheta_0 = (\alpha_0, \nu_0, \beta_0)^T$.
Then $N^{1/2}(\wh{\btheta}  -\btheta_0) \weakto N(0, \uV^{-1} \uSigma \uV^{-1})$ with 	 $\uV$ and $\uSigma$ defined in Supplementary Material. 
The proof of the Theorem 
is given  in Supplementary Material.   

Standard deviations  and 95\% confidence intervals for $\btheta$ can be obtained based on empirical estimates 
of $\uSigma$ and $\uV$ or using a bootstrap resampling procedure that samples controls, incident and prevalent cases with replacement from the respective groups, with fixed sample sizes $n_0, n_1$ and $n_2$ and then fits steps 1 and 2 of the two-step procedure for each bootstrap sample. Confidence intervals can be computed either based on the bootstrap standard deviations assuming normality, or based on the quantiles of the  bootstrap distribution.
 
\section{Simulation study}\label{sec:sim}

We  assessed small sample bias of our two-step  approach and compared it's  efficiency   to several other methods in simulations.

\subsection{Data generation}

We generated data from a retrospective setting.  
 For   controls  we obtained  $n_0$ covariate values from $\uX_0= (X_{01},X_{02})^T\sim N(0,\uSigma^X)$, where $\Sigma^X_{ii}=1$ and $\Sigma^X_{ij}=\Sigma^X_{ji}=0.5$, $i\neq j$.
For incident cases, we used importance sampling to generate $n_1$ covariates  from model (\ref{eq:exptilt}),    $\uX_1= (X_{11}, X_{12})^T \sim f_1$, as follows. We first generated $\tilde n_1$ realizations of $\tilde \uX_1 \sim   N(0,\uSigma^X)$, where $\tilde n_1 >> n_1$. Then we drew a sample of size $n_1$ with replacement where each observation $\tilde \ux_{1,k}$, $k= 1,\dots,n_1$ was sampled with probability
$\exp(   \tilde \ux_{1,k}^T \ubeta) /\sum_{j=1}^{n_1} \exp(   \tilde \ux_{1,j}^T \ubeta)$
which ensures that the resulting sample arises from distribution $f_1$ for  $\ubeta= (\beta_1,\beta_2)^T= (0,0)^T$ and $(1, -1)^T$.  
 
Survival times for incident cases were generated assuming  
$\lambda_0(t)=1$       in model (\ref{eq:prop}), and  thus survival times  $T$ had  an   exponential distribution
 with  $\lambda(t|\uX,\bkappa)= \exp(\uX^T\ugamma)$.

 To obtain covariates for prevalent cases, we  first  generated $\tilde \uX_2$ from $f_1$ as described above.
Given  $\tilde \uX_2$, $T$ was drawn from $S$ with a constant baseline hazard function $\lambda_0(t)=1$,   and only those samples with  $T>A$ were selected, where the backward time $A$ was drawn from a uniform distribution, $A \sim U[0, \xi],$ with $\xi=30$.  This selection procedure tilts the distribution of $\tilde \uX_2$ from $f_1$ to  $f_2$ and thus yields $\uX_2 \sim f_2$.
 Alternatively, one could use  importance sampling with   weights $\tilde w_2(\tilde \ux)= \exp[\tilde \ux^T \ubeta+\log\{\mu(\tilde \ux,\bkappa)\}]$ to generate $\uX_2$, similar  to  the incident cases.

The censoring variables for incident and prevalent cases, $C \sim  U[0, \tau]$ was generated with different  values of $\tau$  to obtain the same amount of censoring among both, incident and prevalent cases.
For incident cases, we set $\tilde T= \min(T, C_1)$, and for prevalent cases, $\tilde T= A+ \min(T-A, C_2)$, i.e. for prevalent cases we censored the forward time, which the difference between the total survival time and the backward time.
We studied the settings of 10\% ($\tau=5$ for incident and $\tau=15$ for prevalent cases), 50\% ($\tau=0.6$ for incident and $\tau=1.5$ for prevalent cases)  and 90\% censoring ($\tau=0.05$ for  incident and  $\tau=0.15$ for  prevalent cases).

The simulation  results in all  tables 
are based on 500 replications for each setting. 


\subsection{Analysis methods}

We compared small sample bias and the efficiency of estimates from our two-step  approach, referred to by
``EM'' in the tables,  to  estimates from several different methods: 
the two-step method  implemented using the truncation-adjusted Cox-partial likelihood  in Section 3 (``Cox''); 
and from ``joint'', i.e. maximizing the full profile likelihood (\ref{eq:fullL}) that also incorporates the prospective follow-up data  assuming a Weibull baseline hazard
jointly for  the survival parameters, $ (\ugamma, \kappa_1, \kappa_2)^T$, and logistic parameters $(\alpha,\nu,\ubeta$).   
  We also show estimates obtained from maximizing the profile likelihood 
(\ref{eq:loglik}) using only backward time information,  termed ``IP-CC'' for incident/prevalent case-control study    \cite{maziarz2019inference}. The backward time distribution   was parameterized using  a Cox model with a Weibull baseline hazard for $S$.

\subsection{Results}

Table \ref{tab1} shows estimates (Est) and empirical standard deviations (SDs) as the   amount of  censoring in the prospective follow-up data of the cases increased  from 10\% to 90\%. The true log-odds ratios  were   $\ubeta=(1, -1)$ and the Cox regression log-hazard ratio (HR) parameters were  $\ugamma=(1, -1)$.

For  $n_0=n_1=n_2=500$  with 10\% and 50\% censoring,    $\wh{\ubeta}$, and  $\wh{\ugamma}$ were unbiased for  all methods, including  all parametric models (joint and IP-CC),
since the survival time was generated from an exponential distribution.
Not surprisingly, the methods that used prospective follow-up time resulted in much small SDs for the log-HR  parameters $\ugamma$ ($SD=0.04$ or $SD=0.05$)  than  the IP-CC method ($SD=0.1$), that only utilizes  the backward time of the prevalent cases.  
The SDs for $\ubeta$ were virtually  the same for all methods ($SD=0.06$ or $SD=0.07$; Table \ref{tab1}). 
  
For 90\% censoring of the prospective case follow-up times,  estimates based on the 
two-step algorithm with the EM  were somewhat biased, with 
$\wh{\ugamma}=(0.78,-0.78)$ and 
 $\wh{\ubeta}=(0.84,-0.85)$. Estimates  $\wh{\ugamma}$ from the two-step algorithm with $\wh{\bkappa}$  from the Cox partial likelihood were unbiased,  but there was  a 12\% bias in  $\wh{\ubeta}=(0.92,-0.92)$ (Table \ref{tab1}).  

The small-sample bias for the two-step procedure with the EM algorithm or the Cox partial likelihood decreased as the sample size increased for either prevalent 
or incident cases (Supplemental Tables 1 and  2, respectively). 
 The bias in $\wh{\ubeta}$ decreased to about 9\% using EM for $n_0= n_1= 500$, $n_2= 1000$ (Supplemental Table 1), but for the Cox method, the bias in $\wh{\ugamma}$ remained at 12\%. For $n_0= 500$, $n_1= 1000$, $n_2= 500$ (Supplemental Table 2), EM based $\wh{\ugamma}$ had a 13\% bias while estimates $\wh{\ugamma}$  from the Cox partial likelihood  were unbiased. The corresponding estimates   $\wh{\ubeta}$ had a 7\% and 10\% bias  for the EM and Cox partial likelihood estimation, respectively.
For both these settings the methods that used a fully parametric specification of the survival function  yielded  unbiased estimates of  all  model parameters. 
   
Supplemental Tables 3, 4 and 5 give  results for $\ubeta=(0,0)$ for different sample sizes. For 
10\%, 50\% and 90\% censoring, all estimates were unbiased for all choices of sample sizes. For 90\% censoring with $n_0= n_1=n_2= 500$, EM-based estimates $\wh{\ubeta}$ had small approximately 7\% bias and  $\wh{\ugamma}$ had 12\% bias. These biases   decreased with increasing sample size for both case groups.

\begin{figure}[ht]
\begin{center}
\includegraphics[scale=0.66]{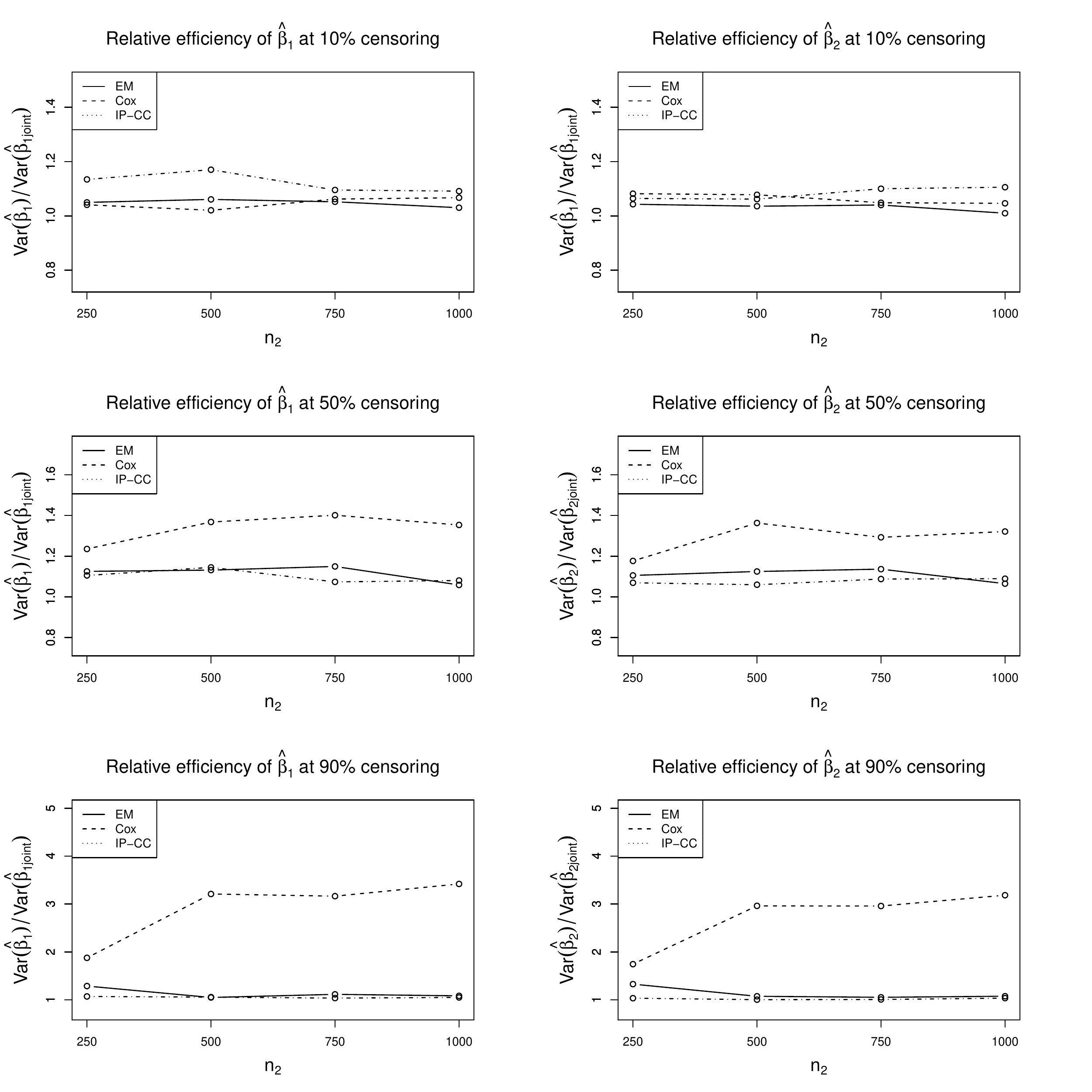}
\caption{Relative efficiency of ($\wh\beta_1$,$\wh\beta_2$) from EM, Cox and IP-CC methods compared to that from joint likelihood method for $n_2$= 250, 500, 750 and 1000 prevalent cases, and under 10\%, 50\% and 90\% censoring when true $(\beta_1,\beta_2) = (1,-1)$.
}\label{replot}
\end{center}
\end{figure}

Figure \ref{replot} shows the relative efficiency, defined as the variance ratio $RE=Var(\wh{\beta})/Var(\wh{\beta}_{joint})$ for $(\wh{\beta}_1, \wh{\beta}_2)$ estimated  using the    EM,   Cox and IP-CC methods compared to the joint likelihood approach
 for  different amounts of censoring as the number of prevalent cases,   $n_2$,   increased from 250 to 1000 (in increments of 250). For all settings in the Figure we used   $n_0=n_1=500$ controls and incident cases. 
For 10\% censoring, all two step methods had the same efficiency as the full likelihood estimates, and it was better than that of the  IP-CC estimates for $n_2=250$ and $500$ for $\beta_1$ and for $n_2=750$ and $n_2=1000$ for $\beta_2$. 
When the amount of censoring increased, the efficiency of the two-step estimator with the Cox model was noticeably worse than the other methods for both components of $\ubeta$. E.g., for 50\% censoring and $n_2=500$, 
the two-step algorithm with the Cox partial likelihood estimates had $RE=1.4$, and for $n_2=750$, $RE= 1.4$ for $\beta_1$ and $1.3$ for $\beta_2$. For 90\% censoring and $n_2= 500$ and $750$, $RE$ for $\beta_1$ was greater than $3$, and for $\beta_2$, $RE=3$. However, the two--step method with the EM estimators exhibited no loss of efficiency compared to the joint likelihood estimation, with all $RE$s around one.
The  IP-CC estimates of
 $\ubeta$, that rely on parametric assumptions for the survival model had similar $RE$s as the two-step EM.

To assess the robustness  of the methods when the underlying baseline did not have a monotone structure  we generated data using two different step-functions for $\lambda_0(t)$ in (\ref{eq:prop}) 
on the intervals  $I_1=[0,7]; I_2= (7,14]; I_3 =  (14, 21]; I_4 = (21, 30]$.  The first hazard function had values $ \lambda_0(t)= 10^{-4},    10^{-5}, 2\times 10^{-4},      0.5\times 10^{-4},$ 
 and the second one   had values    $ \lambda_0(t)= 10^{-5},     2.0\times 10^{-4}, 
10^{-5},      2.0\times 10^{-4},$ for $t \in I_k, k=1,\ldots,4$, respectively. Estimates   $\wh{\ubeta}$ from the EM method were unbiased and comparable to the competing methods. Under both of these baselines, under 90\% censoring, $\wh{\ugamma}$ from  the EM method had a 21\% bias. However,    $\wh{\ubeta}$ was unbiased (Supplemental Tables 6 and 7). 

\section{Data example}\label{sec:data}

We   analyzed data from a case-control study conducted within the USRTS to assess associations of SNPs in candidate genes with risk of breast cancer  \cite{bhatti2008breast}. 
The USRTS, initiated in 1982 by the National Cancer Institute and
other institutions, enrolled $146 022$  radiologic technologists to study 
health effects from low-dose
 occupational radiation exposure.
 Information on participants' characteristics, exposures
and prior health outcomes was collected via several surveys conducted between 1984 and 2014, and blood sample collection  began in 1999.  

 The breast cancer case-control study used information from the first two surveys, conducted  1984-1989 and 1993-1998.  Women who 
answered both surveys and were diagnosed with a breast cancer between the two surveys were considered incident cases and women who answered only one survey and reported  a prior breast cancer diagnosis were considered prevalent cases.  
  All cases  with blood samples  for genetic analysis were included in the study.
 We analyzed data on  711 controls, 386 incident cases, and 227 prevalent cases,  
with  follow-up information on the cases  through December 2008,  available   through regular linkage with the National Death Index. Only 49  breast cancer cases  died during follow-up, corresponding to  92\% censoring. 
 
We modeled the survival distribution using Cox proportional hazards regression (\ref{eq:prop}) with either an unspecified or a Weibull baseline hazard function.
Covariates included in the relative risk model for the survival distribution were:  genotype for the SNP  rs2981582 (1 if TC/TT, 0 if CC); age at breast cancer diagnosis in five categories ($\leq 22$, (22, 40], (40, 50], (50, 55], $> 55$); the year when the woman started working as a  radiation technologist (1 if $\leq  1955, 0$ if $>1955)$, and history of heart disease (yes/no). The logistic model included the following covariates: genotype for all three SNPs: rs2981582; rs889312 (1 if CA/CC, 0 if AA); and rs13281615 (1 if GG/GA, 0 if AA); age at selection or diagnosis;  year first worked; family history of breast cancer (yes/no), BMI during a woman's 20s in three categories ($\leq 20$, $(20, 25]$, $>25$); alcohol consumption (1 if $\geq 7$ drinks per week, 0 otherwise) and an age-BMI interaction, coded as the BMI in a woman's 20s for those selected or diagnosed at or after age 50 and 0 otherwise.

As in the simulations, we fit the two-step method with the truncation adjusted Cox partial likelihood and the EM algorithm, and compared the results  to those obtained from two models with fully specified parametric survival distributions (Cox model with Weibull baseline hazard), the full joint likelihood (\ref{eq:fullL})  and the IP-CC method using  backward time $A$  of the prevalent cases.
We estimated the standard deviations (SDs) using a bootstrap with 500  samples, where we re-sampled controls, incident and prevalent cases with replacement from within each group. 
 
The log-HR estimates and their SDs from the Cox partial likelihood and the EM method were very similar.  The only statistically significant association with survival was seen for ``age at diagnosis'', with women diagnosed at older ages being at higher risk of death. 
 
Estimates of the survival model obtained by maximizing the profile-likelihood (\ref{eq:fullL}) 
 that also includes the prospective follow-up information assuming a Weibull baseline hazard function yielded very similar estimates and SDs as the two-step methods, with the exception of ``year first worked'', which had a larger, statistically significant  log-HR=$-1.03$ than any of the two-step methods. 
The same statistically significant and larger log-HR was also obtained from  the
IP-CC method, that only uses the backward time information. 
While the log-HR estimates were similar across all methods, 
the estimates of the parameters of the Weibull baseline hazard differed noticeably between the IP-CC  method and the methods that used prospective follow-up data. They  were $\wh{ \kappa}_1=3.67$ and $\wh{\kappa}_2=77.01$ for the two-step and   $\wh{\kappa}_1=4.87$ and $\wh{\kappa}_2=47.14$ for the full profile-likelihood   and much lower, $\wh{\kappa}_1=1.46$ and   $\wh{\kappa}_2=10.98$, for the IP-CC method.

 Estimates of the log-odds ratios $\ubeta$, including estimates for the three SNPs, the main exposure, were virtually identical for all methods, with the exception of year first worked, for which    $\wh{\beta}=-0.31 (SD=0.15)$
 for the IP-CC method,  but all other methods  had estimates very close to zero, e.g.   $\wh{\beta}=-0.01 (SD=0.15)$ based on two-step procedure with the EM algorithm.   

\section{Discussion}
\label{sec:disc}

In this paper we proposed and studied a two-step semi-parametric method to incorporate prospective follow-up information into  estimating log-odds ratios for association with disease incidence for case-control studies that include  prevalent cases in addition to or instead of incident cases.

While many authors addressed the issue of length-bias when estimating survival parameters from  a prevalent cohort  \cite[e.g.][]{zhu17a}, the literature on using prevalent cases when samples are ascertained cross-sectionally is limited.
 Begg and Gray \cite{begg84a} adjusted for survival bias when comparing prevalent cases to controls to estimate incidence odds ratios, using a method of moments approach, but did not use any follow-up information.  
 Maziarz et al \cite{maziarz2019inference} proposed a fully efficient method, the IP-CC approach,  to estimate associations of an exposure with disease incidence when a case-control study includes both, incident cases and prevalent cases,  but needed to  model  the backward time for the prevalent cases fully parametrically as  only cross-sectional information on the cases was used.
 
Here, we relax the parametric assumptions and model the  survival distribution semi-parametrically,  using a Cox proportional hazards model. 
To further improve  efficiency of the  estimates of the   model, we also extended the EM algorithm proposed by  Qin et al \cite{qin2011maximum} and Liu et al \cite{ liu2016semiparametric} to accommodate incident and prevalent cases.  
Combining the  survival distribution estimated with the EM algorithm  with the profile-log-likelihood for the two case groups and the controls in a two-step fashion  yielded 
estimates of the log-odds ratio parameters that  were as efficient  as those from jointly maximizing the profile-log-likelihood and  the survival data under a parametrically specified  survival distribution for most settings we studied in simulations. 
Only when there were very large amounts of censoring (90\%) did we observe biases for the EM based estimates for sample sizes of $500$ incident and prevalent cases and 500 controls, that disappeared with larger sample sizes, and thus more deaths after diagnosis. 

  We have the following explanation  for  the bias in the model estimates when there is a very high percentage of  censoring.   As the  density  of the backward time $A$ is $f_A(a)=S(a)/\mu, a>0,$  $f_A(0)=S(0)/\mu=1/\mu$. Thus $\mu$ is the inverse of $f_A(0+)$ (omitting 
$\uX$ for simplicity).
 Woodroofe and Sun  \cite{woodroofe93}  noticed that in general it is not possible to consistently 
estimate the  value   of a non-increasing density at $0+$ non-parametrically and  therefore   $\mu$ is not estimable based on the backward time only. 
However, when the proportion of  censored prospective observations is high, most of the information on $\mu$  comes from the   backward times, and it is therefore not surprising that 
$\wh{\mu}$ and as a consequence estimates of the log-odds ratios  exhibit some    bias.

Our motivating example, the analysis of incident and prevalent breast cancer cases and controls in the USRTS  
 to assess the impact of SNPs on breast cancer risk,  also highlights that when there is a large amount of censoring, adding prospective follow-up information does not  improve efficiency of the association estimates much.
However, if prospective  information is readily available it should be incorporated into the analysis to lessen the reliance on model assumptions that are needed for fitting  the IP-CC method with only cross-sectional information.   

However, when the amount of censoring was limited (around 50\%), the two-step procedure with the EM algorithm was unbiased and much more efficient than estimating the parameters of the survival distribution  using the Cox partial likelihood. In simulations the efficiency of the EM based two-step estimates of the log-odds ratio parameters were as efficient as those obtained from a joint likelihood under parametric assumptions on $S$.
This gain in efficiency comes from the EM also utilizing information on the support of the backward time, $A$. While we assume that $A$ had a uniform distribution, which holds true when the disease process is stationary in the population,  the EM algorithm and our two-step procedure can be implemented using  any  parametrically specified  distribution for $A$   \cite{liu2016semiparametric}. Efficiency gains in log-odds ratio estimates gleaned from the survival information may be lessened  if many parameters in the distribution of $A$ need to be estimated, however. 
Another practically appealing aspect of the two-step procedure with the EM estimation is that it is very easy to  implement using standard survival software and closed form expressions for the baseline hazard function estimates,  
and allows incorporation of all available information on incident and prevalent cases sampled into a case control study.

In summary,  when case-control studies include  prevalent cases 
using  additional  follow-up information on cases is recommended to improve efficiency of estimates of association.



\section*{Acknowledgments}

Data used in this paper are subject to third party restrictions.  The authors thank 
 Jerry Reid,  
 Diane Kampa,  Allison Iwan,  
 Jeremy Miller,  
and the radiologic technologists who participated. This work utilized the computational resources of the NIH HPC Biowulf cluster (http://hpc.nih.gov).

\bibliographystyle{plain}
\bibliography{draft-October5}

\begin{thebibliography}{10}

\bibitem{aalen78}
O.~O. Aalen.
\newblock Nonparametric inference for a family of counting processes.
\newblock {\em Annals of Statistics}, 6:701--726, 1978.

\bibitem{begg84a}
C.~B. Begg and R.~Gray.
\newblock Calculation of polychotomous logistic regression parameters using
  individualized regressions.
\newblock {\em Biometrika}, 71(1):11--18, 1984.

\bibitem{bhatti2008breast}
P.~Bhatti, M.~M. Doody, B.~H. Alexander, J.~Yuenger, S.~L. Simon, R.~M.
  Weinstock, M.~Rosenstein, M.~Stovall, M.~Abend, D.~L. Preston, et~al.
\newblock Breast cancer risk polymorphisms and interaction with ionizing
  radiation among us radiologic technologists.
\newblock {\em Cancer Epidemiology and Prevention Biomarkers},
  17(8):2007--2011, 2008.

\bibitem{breslow72}
N.~Breslow.
\newblock Discussion of paper by d.r. cox.
\newblock {\em Journal of the Royal Statistical Society}, 34:216--217, 1972.

\bibitem{cox72}
D.~R. Cox.
\newblock Regression models and life-tables.
\newblock {\em Journal of the Royal Statistical Society Series B-Statistical
  Methodology}, 34(2):187--202, 1972.

\bibitem{cox75}
D.~R. Cox.
\newblock Partial likelihood.
\newblock {\em Biometrika}, 62(2):269--276, 1975.

\bibitem{liu2016semiparametric}
H.~Liu, J.~Ning, J.~Qin, and Y.~Shen.
\newblock Semiparametric maximum likelihood inference for truncated or
  biased-sampling data.
\newblock {\em Statistica Sinica}, 26:1087--1115, 2016.

\bibitem{maziarz2019inference}
M.~Maziarz, Y.~Liu, J.~Qin, and R.~M. Pfeiffer.
\newblock Inference for case-control studies with incident and prevalent cases.
\newblock {\em Biometrics}, 75(3):842--852, 2019.

\bibitem{newey94}
W.~K. Newey and D.~McFadden.
\newblock Chapter 36 large sample estimation and hypothesis testing.
\newblock {\em Handbook of Econometrics}, 4:2111--2245, 1994.

\bibitem{pfeiffer2017absolute}
R.~M. Pfeiffer and M.~H. Gail.
\newblock {\em Absolute risk: Methods and applications in clinical management
  and public health}.
\newblock 2017.
\newblock cited By 10.

\bibitem{qin1998inferences}
J.~Qin.
\newblock Inferences for case-control and semiparametric two-sample density
  ratio models.
\newblock {\em Biometrika}, 85(3):619--630, 1998.

\bibitem{qin2011maximum}
J.~Qin, J.~Ning, H.~Liu, and Y.~Shen.
\newblock Maximum likelihood estimations and em algorithms with length-biased
  data.
\newblock {\em Journal of the American Statistical Association},
  106(496):1434--1449, 2011.

\bibitem{reid85}
N.~Reid and H.~Crepeau.
\newblock Influence functions for proportional hazards regression.
\newblock {\em Biometrika}, 72(1):1--9, 1985.

\bibitem{schlesselman82}
J.~J. Schlesselman.
\newblock {\em Case-control studies: design, conduct, analysis}.
\newblock Oxford University Press, 1982.

\bibitem{sun2018missing}
Y.~Sun, J.~Qin, C.-Y. Huang, et~al.
\newblock Missing information principle: A unified approach for general
  truncated and censored survival data problems.
\newblock {\em Statistical Science}, 33(2):261--276, 2018.

\bibitem{van00}
A.~W. van~der Vaart.
\newblock {\em Asymptotic Statistics}.
\newblock Cambridge University Press, 2000.

\bibitem{woodroofe93}
M.~Woodroofe and J.~Y. Sun.
\newblock A penalized maximum likelihood estimator of $f(0+)$ when $f$ is
  non-increasing.
\newblock {\em Statistica Sinica}, 3:501--515, 1993.

\bibitem{zhu17a}
H.~Zhu, J.~Ning, Y.~Shen, and J.~Qin.
\newblock Semiparametric density ratio modeling of survival data from a
  prevalent cohort.
\newblock {\em Biostatistics}, 18(1):62--75, 2017.

\end{thebibliography}

\begin{table}
\centering
\caption{Estimates (Est) and empirical standard deviations (SDs) based on 500 replications for $n_0=500, n_1=500, n_2 = 500$. In the population, $(X_1, X_2)^T$ are multivariate normally distributed with mean $(0,0)$, $Var(X_1)=Var(X_2)=1$ and  $Cov(X_1,X_2)=0.5$.  
}\label{tab1}
\begin{tabular}{ll rr rr rr}
 \hline
\multicolumn{1}{c}{Method}& & \multicolumn{1}{c}{$\beta_1$= 1} & \multicolumn{1}{c}{$\beta_2$= -1} & \multicolumn{1}{c}{$\gamma_1$= 1} & \multicolumn{1}{c}{$\gamma_2$= -1} & \multicolumn{1}{c}{$k_1$= 1} & \multicolumn{1}{c}{$k_2$= 1} \\ 
		\cmidrule{1-8}
	& & \multicolumn{6}{c}{10\% censoring}\\
\cmidrule{2-8}
 Two-step &Est (EM) &  1.00 & -1.01 & 1.03 & -1.03 &  &  \\ 
&  SD (EM) &  0.07 & 0.07 & 0.04 & 0.04 &  &  \\ 
&Est (Cox) &  1.00 & -1.00 & 1.00 & -1.00 &  &  \\ 
&  SD (Cox) &  0.06 & 0.07 & 0.05 & 0.05 &  &  \\ 
  Likelihood &  Est (joint) &  1.00 & -1.00 & 1.00 & -1.01 & 1.00 & 1.00 \\ 
&  SD (joint) &  0.06 & 0.07 & 0.04 & 0.04 & 0.03 & 0.04 \\ 
&  Est (IP-CC) & 1.00 & -1.01 & 1.02 & -1.02 & 1.02 & 1.01 \\ 
&  SD (IP-CC) & 0.07 & 0.07 & 0.10 & 0.10 & 0.09 & 0.13 \\ 
\cmidrule{1-8}
		& & \multicolumn{6}{c}{50\% censoring}\\
\cmidrule{2-8}
Two-step  &Est (EM) &  1.03 & -1.04 & 0.99 & -0.99 &  &  \\ 
&  SD (EM) &  0.07 & 0.07 & 0.06 & 0.06 &  &  \\ 
&Est (Cox) &  1.00 & -1.00 & 1.01 & -1.00 &  &  \\ 
&  SD (Cox) & 0.07 & 0.08 & 0.06 & 0.06 &  &  \\ 
  Likelihood &  Est (joint) &  1.00 & -1.00 & 1.01 & -1.01 & 1.01 & 1.01 \\ 
&  SD (joint) &  0.06 & 0.07 & 0.05 & 0.05 & 0.03 & 0.04 \\ 
&  Est (IP-CC) &  1.00 & -1.01 & 1.02 & -1.03 & 1.01 & 1.01 \\ 
&  SD (IP-CC) &  0.07 & 0.07 & 0.10 & 0.10 & 0.09 & 0.13 \\ 
\cmidrule{1-8}
 	     & & \multicolumn{6}{c}{90\% censoring}\\ 
\cmidrule{2-8}
 Two-step &Est (EM) &  0.84 & -0.85 & 0.78 & -0.78 &  &  \\ 
&  SD (EM) &  0.07 & 0.07 & 0.07 & 0.07 &  &  \\  
&Est (Cox) &  0.92 & -0.92 & 1.02 & -1.01 &  &  \\ 
&  SD (Cox) &  0.12 & 0.12 & 0.15 & 0.13 &  &  \\
  Likelihood & Est (joint) &  1.00 & -1.01 & 1.02 & -1.02 & 1.01 & 1.01 \\ 
&  SD (joint) &  0.07 & 0.07 & 0.06 & 0.06 & 0.04 & 0.06 \\ 
&  Est (IP-CC) &  1.00 & -1.01 & 1.02 & -1.03 & 1.01 & 1.01 \\ 
&  SD (IP-CC) &  0.07 & 0.07 & 0.10 & 0.10 & 0.09 & 0.13 \\ 
\hline
\end{tabular}
\end{table} 

\clearpage


\begin{table}
\centering
\caption{Estimates (Est) and bootstrap standard deviations (SD)  based on 500 bootstrap samples in parenthesis.}
\label{tabdata}
\begin{tabular}{l r r r r}
\hline
&\multicolumn{2}{c}{Two-step Estimation }&\multicolumn{2}{c}{Likelihood }\\
Variable&  \multicolumn{1}{c}{EM} & \multicolumn{1}{c}{Cox} & \multicolumn{1}{c}{Joint} & \multicolumn{1}{c}{IP-CC} \\
                  \cmidrule{2-5}
\multicolumn{1}{c}{}&\multicolumn{4}{c}{Log-hazard ratios from Cox proportional hazards model}\\
                  \cmidrule{2-5}
 rs2981582           &  0.22 (0.25) & 0.30 (0.33) &  0.21 (0.25) & -0.02 (0.19) \\ 
 Age at diagnosis   & 0.65 (0.10) & 0.76 (0.16) &  0.80 (0.10) & 0.40 (0.08) \\ 
Year first worked   &  -0.45 (0.24) & -0.23 (0.32) &  -1.03 (0.26) & -1.27 (0.22) \\ 
  History of            &   0.62 (0.30) & 0.75 (0.35) &  0.57 (0.32) & -0.20 (0.31) \\ 
heart disease         &    &    &   &  \\    

\cmidrule{2-5}
&&&\multicolumn{2}{c}{Parameters  of  the Weibull baseline }\\ 

\cmidrule{2-5}
  $k_1$ &    &     & 4.87 (0.36) & 1.46 (0.21) \\ 
  $k_2$ &    &     & 47.14 (4.35) & 10.98 (0.35)\\

                  \cmidrule{2-5}
\multicolumn{1}{c}{}&\multicolumn{4}{c}{Log-odds ratios from logistic model}\\
                  \cmidrule{2-5}
   rs2981582                         &  0.11 (0.12) & 0.11 (0.11) &  0.11 (0.12) & 0.09 (0.12) \\ 
   rs889312                           &  0.24 (0.11) & 0.23 (0.11) &  0.24 (0.11) & 0.24 (0.11) \\ 
   rs13281615                       & 0.29 (0.12) & 0.29 (0.12) &  0.29 (0.12) & 0.30 (0.12) \\ 
  Age at diagnosis/selection    &  -0.02 (0.08) & -0.04 (0.08) & 0.00 (0.08) & 0.04 (0.08) \\ 
 Year first worked                  &  -0.01 (0.15) & 0.02 (0.15) &  -0.05 (0.14) & -0.31 (0.15) \\ 
   Family history                    &  0.51 (0.15) &  0.51 (0.15) & 0.51 (0.15) & 0.51 (0.15) \\ 
   BMI                                  &  -0.33 (0.10) & -0.33 (0.10) &  -0.33 (0.10) & -0.33 (0.10) \\ 
   Age-BMI int.                      &   0.01 (0.01) & 0.01 (0.01) & 0.01 (0.01) & 0.01 (0.01) \\ 
  7+ alcoholic drinks/week     &  0.41 (0.30) & 0.41 (0.30) &  0.41 (0.30) & 0.39 (0.30) \\                 
\hline
\end{tabular}
\end{table}

\clearpage

\noindent{\Large{\bf Supplementary material for: Incorporating survival data to case-control studies with incident and prevalent case}}

\section{Consistency of $(\bkappa, \zeta)$ estimated using the EM method}

As in Liu et al  \cite{liu2016semiparametric}, we allow $A$ to arise from  any known parametric distribution with density $h_\zeta(a)$  and distribution function $H_\zeta(a)$ with parameters $\zeta$, but later use  $H_{\zeta}(a)= U[0, \zeta]$. The  likelihood for the survival data from  incident and prevalent cases is  
\be
L_S(\bkappa,\zeta)= \prod^{n_1}_{i=1}g^{\delta_i}(Y_i|\ux_i,\bkappa) S^{1-\delta_i}(Y_i|\ux_i,\bkappa) \prod^{n_2}_{i=1}\frac{g^{\delta_i}(Y_i|\ux_i,\bkappa) S^{1-\delta_i}(Y_i|\ux_i,\bkappa)}{\int^\xi_0 S(u|\ux_i,\bkappa)h_\zeta(u)du}h_{\zeta}(A_i).
\ee
 The corresponding  log-likelihood under a Cox proportional hazards model for $S$ is
\be
l_S(\bkappa,\zeta) &=& \sum^{n_1}_{i=1}\Big[\sum^K_{k=1}[\delta_i\{\log\lambda(t_k)+\ux^T_i\ugamma\}-\Lambda(t_k)\exp(\ux^T_i\ugamma)]I(Y_i=t_k)\Big] +\sum^{n_2}_{i=1}\Big[\sum^K_{k=1}[\delta_i\{\log\lambda(t_k)\nonumber\\
 && +\ux^T_i\ugamma\}-\Lambda(t_k)\exp(\ux^T_i\ugamma)]I(Y_i=t_k)\Big]+\log h_\zeta(A_i) -\log\int S(u|\ux_i,\bkappa)h_\zeta(u)du.
\label{sup:logl}
\ee
Define $N^I_i(t)= I(Y_i\leq t)\delta_i$, $N^P_i(t)= I(A_i<Y_i\leq t)\delta_i$, $M^I_i(t)= I(Y_i\geq t)$ and $M^P_i(t)= I(Y_i\geq t)I(Y_i>A_i)$. The score functions for $l_S$ are
\bse
U_{1n}&=& \nabla_{\zeta}l_S(\bkappa,\zeta)= \sum^{n_2}_{i=1}\Big\{\frac{\dot h_\zeta(A_i)}{h_\zeta(A_i)}I(Y_i>A_i)-\frac{\int^\xi_0 S(u|\ux_i,\bkappa)\dot h_\zeta(u)du}{\int^\xi_0 S(u|\ux_i,\bkappa)h_\zeta(u)du}\Big\}\\
U_{2n}&=& \nabla_{\ugamma}l_S(\bkappa,\zeta)= \sum^{n_1}_{i=1}\Big[\int^\xi_0 dN^I_i(u)-\int^\xi_0 M^I_i(u)\exp(\ux^T_i\ugamma)d\Lambda(u)\Big]\\
&& + \sum^{n_2}_{i=1}\Big[\int^\xi_0 \Big\{ dN^P_i(u)- (M^P_i(u)-\frac{\int^\xi_u S(v|\ux_i,\bkappa)h_\zeta(v)dv}{\int^\xi_0 S(v|\ux_i,\bkappa)h_\zeta(v)dv})\exp(\ux^T_i\ugamma)d\Lambda(u)\Big\}\Big]\\
U_{t, 3n}&=& \nabla_{\Lambda}l_S(\bkappa,\zeta)= \sum^{n_1}_{i=1}\Big[\int^t_0 dN^I_i(u)-\int^t_0 M^I_i(u)\exp(\ux^T_i\ugamma)d\Lambda(u)\Big]\\
&& + \sum^{n_2}_{i=1}\Big[\int^t_0 \Big\{ dN^P_i(u)- (M^P_i(u)-\frac{\int^\xi_u S(v|\ux_i,\bkappa)h_\zeta(v)dv}{\int^\xi_0 S(v|\ux_i,\bkappa)h_\zeta(v)dv})\exp(\ux^T_i\ugamma)d\Lambda(u)\Big\}\Big]
\ese
where $\dot h_\zeta(\cdot)$ is the partial derivative of $h_\zeta(\cdot)$ with respect to $\zeta$. Next we show that the score equations have mean zero. To show $E_{(T,A)}\{U_{1n}(\zeta, \ugamma, \Lambda)\}= 0$, consider
\bse
E_{(T,A)}\Big[\frac{\dot h_\zeta(A)}{h_\zeta(A)}I(T>A)\Big]= \frac{\int^\xi_0 \dot h_\zeta(a)\int^\xi_0 I(T>A)g(t|X)dt da}{\int^\xi_0 S(u|X,\bkappa)h_\zeta(u)du}
= \frac{\int^\xi_0 S(u|X,\bkappa)\dot h_\zeta(u)du}{\int^\xi_0 S(u|X,\bkappa)h_\zeta(u)du}.
  \ese
Of the remaining score equations, it is enough to show that $E_{(T,A)}\{U_{t, 3n}(\zeta, \ugamma, \Lambda)\}= 0$ since $U_{2n}(\zeta, \ugamma, \Lambda)$ only differs from $U_{t, 3n}(\zeta, \ugamma, \Lambda)$ by an extra $X$ term along with a fixed upper limit, $\xi$, for the integral. We  show the mean of $U_{t, 3n}(\zeta, \ugamma, \Lambda)$ is zero by considering the incident and prevalent case contributions separately. For the incident cases,
\begin{multline} 
E_T[dN^I_i(t)]= \pr[dN^I_i(t)= 1] = \pr[t- < Y_i \leq t, \delta_i= 1]\\
= \pr[t-<Y_i\leq t, \delta_i=1|Y_i>t-]\pr(Y_i>t-) 
= \lambda^*(t)dt E(M^I_i(t))
\end{multline} 
where $\lambda^*(t)$ is the general hazard function at time $t$. Thus  $E_T[dN^I_i(t)]= E[M^I_i(t)]\exp(\ux^T_i\ugamma)d\Lambda(t)$.

Next we show that part of $l_{S}$ corresponding to  the prevalent cases has mean zero. Define $\bar H(a)= 1-H(a)$. Recall that  the latent data for the $i$th prevalent case are $O_i^*= \{(T^*_{ij}, A^*_{ij})\}$, $i= 1,\dots,n_2$, $j= 1,\dots,m_i$ where $T^*_{ij}<A^*_{ij}$. Define $\rho= \pr(T<A)= \int^\xi_0\bar H(u)dG(u)$. Then, $(T^*, A^*)\sim (T, A)|T<A = g(t)h(a)/\rho$, $t<a$. This gives the marginal distribution,
\bse
T^*\sim g^*(t)&=&\frac{1}{\rho}\int^\xi_t g(t)h(a)da = \frac{g(t)\bar H(t)}{\rho}.
\ese
Following Sun et al \cite{sun2018missing}, for left-truncated but uncensored data with a known truncation time distribution, $H_{\zeta}$, we define the following stochastic processes for observed and latent observations respectively
\bse
\tilde M_{i}&=& N_i(t) - \int^t_0 I(T_{i}\geq u)d\Lambda(u)\\
\tilde M^*_{ij}&=& N^*_{ij}(t) - \int^t_0 I(T^*_{ij}\geq u)d\Lambda(u)
\ese
where $N_i(t)= I(T_{i}\leq t)$ and $N^*_{ij}(t)= I(T^*_{ij}\leq t)$. The unbiasedness of the score equations with complete data implies that $d\tilde M^H_i(t)= d\tilde M_i(t) + \sum^{m_i}_{j=1}d\tilde M^*_{ij}(t)$ has zero mean. Since $T^*_{ij}$ are not observed, we replace $\sum^{m_i}_{j=1}d\tilde M^*_{ij}(t)$ by its expected value. Now, 
\bse
E\{d\tilde M^*_{ij}(t)\}= E\{d\tilde N^*_{ij}(t)\} - d\Lambda(t)E\{I(T^*_{ij}\geq t)\}.
\ese
The expectations of the terms on the right hand side are
\bse
E\{d\tilde N^*_{ij}(t)\}&=& \pr(t-<T^*_{ij}\leq t)= g^*(t),\\
E\{I(T^*_{ij}\geq t)\}&=& \pr(T^*_{ij}\geq t)= \frac{1}{\rho}\int^\xi_t \bar H(t)dG(t).
\ese
Since   the $i$th observed prevalent case can be viewed as the first ``success'' after $m_i$ failures, the random integer $m_i$ has a negative binomial distribution with parameters $(1,\rho)$. Thus $E(m_i)= \rho/(1-\rho)$. Then,
\bse
E\{\sum^{m_i}_{j=1}d\tilde M^*_{ij}(t)\}&=& E(m_i)E\{\sum^{m_i}_{j=1}d\tilde M^*_{ij}(t)|m_i\}\\
&=&\frac{1}{1-\rho}\{\bar H(t)dG(t) - d\Lambda(t)\int^\xi_t \bar H(t)dG(t)\}.
\ese
Note that the denominator can be written as 
\be\label{eq:prtga}
1-\rho = \pr(T>A)= \int^\xi_0\bar G(a)dH(a)= \int^\xi_0 H(u)dG(u).
\ee
It follows that
\bse
d\tilde M^H_i(t)= \{d\tilde N_i(t) - I(Y_i\geq t)d\Lambda(t)\}+\frac{\bar H(t)dG(t) - d\Lambda(t)\int^\xi_t \bar H(t)dG(t)}{\int^\xi_0 H(u)dG(u)}.
\ese
For left-truncated, right-censored data with known truncation distribution, $H$, we replace $d\tilde N_i(t)$ and $I(Y_i\geq t)$ by their conditional expectation in the above expression.
\bse
d\tilde M^H_i(t)&=& [E\{d\tilde N_i(t)|Y_i, \delta_i\} - E\{I(Y_i\geq t)|Y_i, \delta_i\}d\Lambda(t)]+\frac{\bar H(t)dG(t) - d\Lambda(t)\int^\xi_t \bar H(t)dG(t)}{\int^\xi_0 H(u)dG(u)}\\
&=& \{dN^P_i(t) - M^P_i(t)d\Lambda(t)\}+\frac{\bar H(t)dG(t) - d\Lambda(t)\int^\xi_t \bar H(t)dG(t)}{\int^\xi_0 H(u)dG(u)}.
\ese
Comparing $d\tilde M^H_i(t)$ with the prevalent part of $U_{t,3n}(\zeta, \ugamma, \Lambda)$, the only remaining step is to show that the fractions in these two expressions are equal. Note that the denominators have already been shown to be equal in (\ref{eq:prtga}).  As $dG(t)/d\Lambda(t)= g(t)/\lambda(t)= S(t)= \int^\xi_t dG(u)$, 
\begin{multline*}
\bar H(t)dG(t) - d\Lambda(t)\int^\xi_t\bar H(u)dG(u)
= d\Lambda(t)\int^\xi_t \{\bar H(t)- \bar H(u)\}dG(u) \\
= d\Lambda(t)\int^\xi_t \pr(t<A<u)dG(u)
= d\Lambda(t)\int^\xi_t \int^u_t dH(s) dG(u) 
= d\Lambda(t)\int^\xi_t \bar G(s)dH(s).
\end{multline*}
This completes  the proof.

 \section{Asymptotic properties of the two-step estimator}

We briefly  summarize the  two-step approach. Let $\nabla_{\phi}$ denote the differentiation operator with respect to a generic parameter $\phi$.

{\bf {\it Step 1.}  Estimate    $\bkappa$ in $  \lambda(t|\ux,\bkappa)=\lambda_0(t) \exp(\ux^T\ugamma),  $ }
  
\noindent   {\it   Estimating $\bkappa$  with a non-parametric   $\lambda_0(t)$ based on the Cox partial likelihood} 
   
Recall that  we first estimate the  log-hazard ratio parameters  $\ugamma$ using   the score functions of  the standard    partial likelihood  \cite{cox72, cox75}, that accommodate the prevalent cases in the definition of the risks sets, $Z(t)$, which do not monotone decrease with $t$, as  the prevalent cases  are at risk only since  their truncation time,  
\begin{align}
\U_{11}(\ugamma) = \sum_{i=1}^n   \sum_{t \geq 0}\bigg\{ {\ux}_i - \dfrac{\wh  {S}^{(1)}(t;{\ugamma})}{\wh  {S}^{(0)}(t; {\ugamma})}\bigg\}\mathrm{d}N_i(t)=\0,  
\label{eq:u11}
\end{align}
with $\wh {S}^{(r)}(t;\ugamma) =\sum_{j=1}^n Z_{j}(t) \exp(\ugamma^\text{T} {\ux}_j){{\ux}_j}^{\otimes r},$ where for a column vector $a, a^{\otimes 0}=1, a^{\otimes 1}=a $ and $a^{\otimes 2}=aa^{T}$.    
Given $\widehat{\ugamma}$, the estimating equations for $\lambda_0(t)$ that lead to the Breslow estimate 
of $\Lambda_{0}$  are
\begin{align}
 \U_{12}\{\lambda_0(t)\mid{\widehat{\ugamma}}\}= \sum_{i=1}^n 
\big\{\mathrm{d}N_i(t)-\lambda_0(t) Z_{i}(t)\exp(\widehat{\ugamma}^\text{T} {\ux}_i)\big\}=\0.  
\label{eq:u12}
\end{align}   

{\bf 

{\it Step 2.}  Estimate    $\btheta=(\alpha, \nu, \ubeta)$ treating   $\bkappa$ as known.    }
This is done by maximizing the  pseudo log-likelihood 
\be \label{eq:twostepL}
\ell(\btheta| \hat \bkappa)&=& -\sum_{i=1}^N\log[1+\exp(\alpha+\ux^T_i\ubeta)+\exp\{\nu+\ux^T_i\ubeta+\log\mu(\ux_i,\hat \bkappa)\}] \nonumber \\
&& +\sum_{i=n_0+1}^{n_0+n_1}(\alpha+\ux^T_i\ubeta)+
\sum_{i=n_0+n_1+1}^N (\nu+\ux^T_i\ubeta),   
\ee
where $N=n_0+n_1+n_2$. 
In what follows, we rewrite  $ l(\btheta, \hat \bkappa)$ in (\ref{eq:twostepL}) using   $ w_1(x) = \exp(\alpha + x \beta)$,
$w_2(x) = \exp\{\nu + x \beta +\log \mu(x,\bkappa)\} $
and
$ \eta(x)=1+w_1(x) +w_2(x)$  as 
\be
\ell(\btheta| \hat \bkappa) = -\sum_{i=1}^N\log \eta(x) 
+ \sum_{i=n_0+1}^{n_0+n_1} \log w_1(x) +\sum_{i=n_0+n_1+1}^N 
 \log w_2(x).
 \ee
%
 %
The corresponding score equations  $\U_2 = (U_{12}, U_{22}, U_{23})$ are 
\be
U_{21}
&=&
N^{-1/2} \nabla_{\alpha} \ell(\btheta| \hat \bkappa)
= - N^{-1/2} \left\{ \sum_{i=1}^N\frac{w_1(x_i) }{\eta(x_i)} - n_1 \right\}
= N^{-1/2} \sum_{i=1}^N\left\{ \delta_{1i} - \frac{w_1(x_i)}{\eta(x_i)}\right\},
\nonumber \\
U_{22}
&=& N^{-1/2} \nabla_{\nu} \ell(\btheta| \hat \bkappa)
=-N^{-1/2} \left\{ \sum_{i=1}^N\frac{w_2(x_i) }{\eta(x_i)} - n_2 \right\}
=
N^{-1/2} \sum_{i=1}^N\left\{ \delta_{2i} - \frac{w_2(x_i)}{\eta(x_i)}\right\},\nonumber \\
U_{23}
&=& N^{-1/2} \nabla_{\beta} \ell(\btheta| \hat \bkappa)
=-N^{-1/2}\left\{ \sum_{i=1}^N\frac{w_1(x_i) +w_2(x_i) }{
	\eta(x_i)} x_i +
 \sum_{i=n_0+1}^{N}x_i\right\} \nonumber\\
&=&
N^{-1/2} \sum_{i= 1}^{N}\left\{ \frac{ 1}{ \eta(x_i)} - \delta_{0i} \right\} x_i,
\label{eq:U2}
\ee
  where 
$\delta_{0i} = I(x_i \mbox{   from a control})$, $\delta_{1i} = I(x_i \mbox{   from an incident case})$
and $\delta_{2i} = I(x_i \mbox{  from a prevalence case})$  \cite{maziarz2019inference}.
 

\subsection{Asymptotic distribution of $\wh{\btheta}$ }
\label{sec:parametriclambda}

As shown in Appendix A.2 of Qin et al \cite{qin2011maximum}, the estimate $\wh{\xi}=\max\{R_1 Y_1,\ldots,R_nY_{n}\}$, namely the maximum of all observed survival times in the  prevalent cases, converges to the true parameter $\xi$ at a rate that is faster than $n_2^{-1/2}$.  Therefore we treat $\xi$ as a fixed constant  in what follows, and only consider the estimation of $\bkappa$.

We have two sets of estimating equations that are solved successively. First we solve the estimating equations
 $\U_{1}$     to obtain estimates $\wh{\bkappa}$ and then  $\btheta$ is estimated based on the estimating  equations (\ref{eq:U2}) for  $\U_{2}$. In summary,  we solve 
 \be
\label{eq:paramtricU}
\begin{bmatrix}
\hat \U_{1}( \bkappa)     \\
  \hat \U_2(\btheta| \hat \bkappa)  
\end{bmatrix}
=
\begin{bmatrix} \0   \\
\0\end{bmatrix}.
\ee




{\it Theorem 1:}  When   $\lambda_0$ is estimated non-parametrically, the estimates $\wh{\btheta}$ that solve the system of equations (\ref{eq:paramtricU})  are consistent and asymptotically normally distributed, $
\sqrt{n} (\wh{\btheta} - \btheta_0) \to N(0,\V^{-1} \Sigma \V^{-1})$,
with 
\be \label{vars} \V = - E [ \nabla_{\theta} \U_2(\ux_i, \btheta_0, \bkappa_0)], 
\mbox{ and }  \Sigma=Var[\U_2(\uX,\btheta_0,\bkappa_0)
+ \nabla_{\mu} \U_2(\uX,\btheta_0, \mu) \Delta\{\mu( \bkappa)\}], \ee 
where $\Delta(.)$ denotes the influence function operator based on von Mises expansions  \cite[Ch.20]{van00}.

{\it Proof:} 
Under the assumption of uniform convergence of each estimating equation component $\U_k$ for all $k$ to a corresponding nonrandom function, we have a consistent estimator for each parameter  \cite[Ch.5.2]{van00}. 

The normality result follows from applying   Theorem 8.1  from Newey and McFadden \cite{newey94}, using that 
$N^{-1/2} \sum_{i=1}^N \U_2(\ux_i,\btheta_0,\wh{\bkappa}) \to N(0,\Sigma)$ in probability, 
where $\Sigma=Var[\U_2(\uX,\btheta_0,\bkappa_0)
+ \delta(\uX)]$. 
The term $\delta$ can be computed based on   influence functions, from  
$\nabla_{\mu} \U_2(\uX,\btheta_0, \mu(\wh{\bkappa})) \Delta(\mu(\wh{\bkappa}))$
where 
\bse
\Delta(\mu) = -\int_0^\xi \exp\{-\Lambda(t|\uX,\bkappa)\} \Delta\{\Lambda(t|\uX,\bkappa)\} dt
\ese
with 
\bse
 \Delta\{\Lambda(t|\uX,\bkappa)\} = \exp(\ugamma' \uX) \big[ \Delta\{ \Lambda_0(t)\} + \Lambda_0(t) \Delta(\ugamma)\big].
\ese
Following Reid and Crepeau \cite{reid85}, we give the  influences for $ \ugamma$ and the derive the influence for $\Lambda_0$. 
Let $ {S}^{(r)}(t;\ugamma) =\int  Z_{j}(s) \exp(\ugamma' {\uX}){{\uX}}^{\otimes r} dP(t,N,\uX)$. The measure $P$ refers to the distribution of the counting process and the covariates. 
The influences for $\ugamma$ and $\lambda_0$ are 
 \begin{equation}
 \Delta \lbrace \ugamma \rbrace= \mathcal{E}( \ugamma)^{-1}  \left[ \lbrace \ux - \vec{E}(\ugamma,t) \rbrace   -    \exp (   \ugamma'\ux ) \int    \frac{\lbrace \ux - \vec{E}( \ugamma,s) \rbrace }{{S}^{(0)}(s;\ugamma) }dP(s,N,\uX)
 \right],    
 \label{eq:cox_delta}
\end{equation}  
 \begin{equation}
 \Delta \lbrace   \Lambda_0(t)\rbrace= -\Delta \lbrace  \ugamma\rbrace 
\int_0^t       \vec{E}(\ugamma,t)\lambda_0(s)ds - 
 \left[\exp (  \ugamma'\ux ) \int_0^{\min(t,Y)}       
 \frac{\lambda_0(s) }{{S}^{(0)}(s;\ugamma) }ds -  \frac{I( Y \leq t)}{{S}^{(0)}(Y;\ugamma) }
\right],  
 \label{eq:baseline_delta}
 \end{equation}  
where 
\begin{equation}\vec{E}(\ugamma,t) = \frac{  {S}^{(1)}(t;\ugamma)}{{S}^{(0)}(t;\ugamma) },  
  \quad   \mathcal{E}( \ugamma) =     \int  
  \left[  \frac{  {S}^{(2)}(t;\ugamma)}{{S}^{(0)}(t;\ugamma) } - \vec{E}( {\ugamma},t)^{\otimes 2} \right] dP(t,N,\uX).  
\label{eq:second_der}
\end{equation}

\subsubsection{Empirical estimates of the asymptotic variance}
 
Empirical estimates are obtained by plugging in the Cox partial likelihood estimates $\hat \ugamma$ and $
\wh{\Lambda}_0(t)$ into equations  (\ref{eq:cox_delta}) and (\ref{eq:baseline_delta}), see also   
   \cite[Ch.4.6]{pfeiffer2017absolute}.  

For   $\hat{\ugamma}$, following  \cite{reid85}, we get   
\begin{equation}
 \wh{\Delta}_i \lbrace \widehat{\ugamma}\rbrace= \mathcal{\wh{E}}(\widehat{\ugamma})^{-1} \sum_t \left[ \lbrace \ux_{i} - \vec{E}(\widehat{\ugamma},t) \rbrace \left(  dN_i(t) 
-  z_i(t) \exp ( \widehat{\ugamma}'\ux_{i} )  
 \frac{\sum_{l}   dN_l(t)}{\wh {S}^{(0)}(s;\ugamma)
} \right )
\right],  
 \label{eq:cox_deltaest}
\end{equation} where $\mathcal{E}(\widehat{\ugamma})$ is minus the second partial derivative of the  log-pseudo-likelihood, 
   \begin{equation}    \mathcal{\wh{E}}(\widehat{\ugamma}) =     \sum_t  \left[  \sum_{i}  \ux_{i}{\ux_{i}}' e_{i}(t) - \vec{E}(\widehat{\ugamma},t) \vec{E}(\widehat{\ugamma},t)' \right],  
\label{eq:second_der}
\end{equation}where ${\vec{E}}(\widehat{\ugamma},t)$ is estimated as 
 \begin{equation}\vec{E}(\ugamma,t) =  \frac{ \wh {S}^{(1)}(t;\wh{\ugamma})}{\wh {S}^{(0)}(t;\wh{\ugamma}) }   
  =      \frac{  \sum_{i=1}^{n}  Z_i(t) \exp (\wh {\ugamma}' \ux_i ) \ux_i}{\sum_{i=1}^n  Z_i(t) \exp (\wh{\ugamma}' \ux_i ) }. 
\label{eq:Hbar}
\end{equation}  
 and 
$ e_{i}(t) =  \ux_{i}(t) \exp ( \hat{\ugamma}' \ux_{i} ) / \sum_{i}  z_{i}(t) \exp ( \hat{\ugamma}' \ux_{i} ).$
  The first term inside the summation in equation (\ref{eq:cox_deltaest}), $dN_i(t) \lbrace \ux_{i} - \vec{E}(\hat{\ugamma},t)\rbrace, $  is  non-zero only if person $i$ has an event.


  Estimates of the influences for the cumulative baseline hazard  are obtained using 
$\wh  \Delta_i \lbrace \widehat\Lambda_0(t)\rbrace =   \sum_{s\leq t}\wh  \Delta_i  \lbrace \widehat\lambda_0(s)\rbrace. $
Straight forward calculation yields 
\bse
\wh {\Delta}_{i} \lbrace \hat{\lambda}_{0} (t) \rbrace = \wh {S}^{(0)}(t;\wh {\ugamma}) ^{-1} \left[  dN_{i}(t) - \hat{\lambda}_{0}(t) \wh {\Delta}_{i} \lbrace \wh {S}^{(0)}(t;\wh {\ugamma})  \rbrace \right]
\ese  
with
 $  \wh  {\Delta}_{i} \lbrace \wh {S}^{(0)}(t;\wh{\ugamma})  \rbrace =   z_{i}(t) \exp ( \widehat{\ugamma}' \textbf{x}_i ) 
 + \left[ \sum_{l} \textbf{x}_l  z_{l}(t) \exp (  \widehat{\ugamma}' \textbf{x}_l  ) \right] \Delta_{i} \lbrace \widehat{\ugamma} \rbrace. $

\newpage

\subsection{Explicit expressions for   $\nabla_{\theta} \U_2(\btheta,\Lambda_0)$}

Here we give explicit expressions for $\nabla_{\theta} \U_2(\btheta,\Lambda_0)$, based on  the Supplementary Material of Maziarz et al  \cite{maziarz2019inference}. 
Define $\rho_i = n_i/N$ and let $E_0$ denote the expectation with respect to $dF_0(x)$, the covariate distribution in controls.  
  
\bse
	N^{-1} \nabla_{\alpha\alpha} \ell
	&=& -\frac{1}{N}\sum_{i=1}^N\frac{w_1(x_i) \{ 1+w_2(x_i) \}}
	{\{\eta(x_i)\}^2}\\
	&\rightarrow &
	-\rho_0 \int \frac{w_1(x ) \{ 1+w_2(x ) \}}
	{\eta(x)}dF_0(x) 
\\ 
	N^{-1} \nabla_{\alpha\nu} \ell
	&=& \frac{1}{N}\sum_{i=1}^N\frac{w_1(x_i) w_2(x_i) }
	{\{\eta(x_i)\}^2}\\
	&\rightarrow &
	\rho_0\int \frac{w_1(x ) w_2(x ) }
	{\eta(x)}dF_0(x) 
	\\
	N^{-1} \nabla_{\alpha\beta} \ell
	&=& -\frac{1}{N}\sum_{i=1}^N\frac{w_1(x_i) }{\{\eta(x_i)\}^2} x_i \\
	&\rightarrow &
	-\rho_0 \int \frac{ w_1(x)}{\eta(x)} x dF_0(x)
	\\
	N^{-1} \nabla_{\alpha\gamma} \ell
	&=& \frac{1}{N}\sum_{i=1}^N\frac{w_1(x_i)w_2(x_i)}
	{\{\eta(x_i)\}^2} \nabla_{\gamma} \{ \log \mu(x_i,\gamma_0)\} \\
	&\rightarrow & \rho_0 \int \frac{w_1(x )w_2(x )}
	{\eta(x)} \nabla_{\gamma} \{ \log \mu(x ,\gamma_0)\} dF_0(x) 
\\
 	N^{-1} \nabla_{\nu\nu} \ell
	&=& -\frac{1}{N}\sum_{i=1}^N\frac{w_2(x_i)
		[1+w_1(x_i)]}
	{\{\eta(x_i)\}^2}\\
	&\rightarrow & -\rho_0 \int \frac{ \{1+w_1(x )\} w_2(x )}
	{\eta(x)}dF_0(x) 
	\\
	N^{-1} \nabla_{\nu\beta} \ell
	&=& -\frac{1}{N}\sum_{i=1}^N\frac{x_iw_2(x_i) }
	{\{\eta(x_i)\}^2}
	\rightarrow
	-\rho_0 \int \frac{xw_2(x )}
	{\eta(x)}dF_0(x) 
	\\
\end{eqnarray*}
\begin{eqnarray*}
	N^{-1} \nabla_{\nu\gamma} \ell
	&=& -\frac{1}{N}\sum_{i=1}^N\frac{[1+w_1(x_i)]w_2(x_i) }
	{\{\eta(x_i)\}^2}\nabla_{\gamma} \{ \log \mu(x_i,\gamma_0)\} \\
	&\rightarrow &
	-\rho_0 \int \frac{[1+w_1(x )]w_2(x )}
	{\eta(x)} \nabla_{\gamma} \{ \log \mu(x_i,\gamma_0)\} dF_0(x)\\
	N^{-1} \nabla_{\beta\beta} \ell
	&=& -\frac{1}{N}\sum_{i=1}^N\frac{x_ix_i^\T\{ w_1(x_i)+w_2(x_i)\}
	}
	{\{\eta(x_i)\}^2}\\
	&\rightarrow &
	-\rho_0 \int \frac{ w_1(x )+w_2(x ) }
	{\eta(x) } xx^\T dF_0(x) 
	\\
	N^{-1} \nabla_{\beta\gamma} \ell
	&=& -\frac{1}{N}\sum_{i=1}^N\frac{x_iw_2(x_i)
	}
	{\{\eta(x_i)\}^2}\nabla_{\gamma} \{ \log \mu(x_i,\gamma_0)\}\\
	&\rightarrow & -\rho_0 \int \frac{xw_2(x)}
	{\eta(x)}\nabla_{\gamma} \{ \log \mu(x,\gamma_0)\}dF_0(x) \\ 
	N^{-1} \nabla_{\gamma\gamma} \ell
	&=& -\frac{1}{N}\sum_{i=1}^N\frac{[1+w_1(x_i)]w_2(x_i)
	}
	{\{\eta(x_i)\}^2}
	\nabla_{\gamma} \{ \log \mu(x_i,\gamma_0)\} \nabla_{\gamma^\T} \{ \log \mu(x_i,\gamma_0)\}
	\\
	&&
	- \frac{1}{N} \sum_{i=1}^N\frac{w_2(x_i) }{\eta(x_i)}
	\nabla_{\gamma\gamma} \{ \log \mu(x_i,\gamma_0) \}
	 \\
	&\rightarrow &  
-\rho_0 \e_0 \left[ \frac{[1+w_1(X )]w_2(X )}
	{\eta(X)}
	[\nabla_{\gamma} \{ \log \mu(X,\gamma_0)\}]^{\otimes 2}
	\right]\\
&&	- \rho_0 \e_0 [ w_2(X)
	\nabla_{\gamma\gamma} \{ \log \mu(X,\gamma_0) \} ]
\end{eqnarray*}


\begin{table}
\centering
\caption{Estimates (Est) and empirical standard deviations (SDs) based on 500 replications for $n_0=500, n_1=500, n_2 = 1000$. In the population, $(X_1, X_2)^T$ are multivariate normally distributed with mean $(0,0)$, $Var(X_1)=Var(X_2)=1$ and  $Cov(X_1,X_2)=0.5$.}
\begin{tabular}{ll rr rr rr rr}
 \hline
\multicolumn{1}{c}{Method}& & \multicolumn{1}{c}{$\alpha$} &\multicolumn{1}{c}{$\nu$} & \multicolumn{1}{c}{$\beta_1$= 1} & \multicolumn{1}{c}{$\beta_2$= -1} & \multicolumn{1}{c}{$\gamma_1$= 1} & \multicolumn{1}{c}{$\gamma_2$= -1} & \multicolumn{1}{c}{$k_1$= 1} & \multicolumn{1}{c}{$k_2$= 1} \\ 
		\cmidrule{1-10}
	& & \multicolumn{8}{c}{10\% censoring}\\
\cmidrule{2-10}
 Two-step & Est (EM) & -0.49 & 0.57 & 0.99 & -0.99 & 1.04 & -1.04 &  &  \\ 
&  SD (EM) & 0.03 & 0.03 & 0.06 & 0.06 & 0.04 & 0.04 &  &  \\ 
&Est (Cox) & -0.50 & 0.69 & 1.00 & -1.00 & 1.00 & -1.00 &  &  \\ 
&  SD (Cox) & 0.03 & 0.04 & 0.06 & 0.06 & 0.04 & 0.04 &  &  \\ 
  Likelihood &  Est (joint) & -0.50 & 0.70 & 1.00 & -1.00 & 1.00 & -1.00 & 1.00 & 1.00 \\ 
&  SD (joint) & 0.03 & 0.03 & 0.06 & 0.06 & 0.03 & 0.03 & 0.02 & 0.04 \\ 
&  Est (IP-CC) & -0.50 & 0.70 & 1.00 & -1.00 & 1.01 & -1.01 & 1.01 & 1.00 \\ 
&  SD (IP-CC) & 0.03 & 0.07 & 0.06 & 0.07 & 0.07 & 0.07 & 0.06 & 0.09 \\ 
\cmidrule{1-10}
		& & \multicolumn{8}{c}{50\% censoring}\\
\cmidrule{2-10}
Two-step  & Est (EM) & -0.53 & 0.47 & 1.04 & -1.04 & 1.04 & -1.04 &  &  \\ 
&  SD (EM) & 0.03 & 0.04 & 0.06 & 0.07 & 0.04 & 0.04 &  &  \\  
&Est (Cox) & -0.50 & 0.68 & 1.00 & -1.00 & 1.00 & -1.00 &  &  \\ 
&  SD (Cox) & 0.04 & 0.05 & 0.07 & 0.07 & 0.06 & 0.05 &  &  \\
  Likelihood &  Est (joint) & -0.50 & 0.69 & 1.00 & -1.00 & 1.00 & -1.00 & 1.00 & 1.00 \\ 
&  SD (joint) & 0.03 & 0.03 & 0.06 & 0.06 & 0.04 & 0.04 & 0.02 & 0.04 \\ 
&  Est (IP-CC) & -0.50 & 0.70 & 1.00 & -1.00 & 1.01 & -1.01 & 1.00 & 1.00 \\ 
&  SD (IP-CC) & 0.03 & 0.07 & 0.06 & 0.07 & 0.07 & 0.07 & 0.06 & 0.09 \\ 
\cmidrule{1-10}
 	     & & \multicolumn{8}{c}{90\% censoring}\\ 
\cmidrule{2-10}
 Two-step& Est (EM) & -0.43 & 0.16 & 0.91 & -0.91 & 0.97 & -0.96 &  &  \\ 
&  SD (EM) & 0.03 & 0.07 & 0.06 & 0.07 & 0.06 & 0.06 &  &  \\ 
&Est (Cox) & -0.48 & 0.60 & 0.95 & -0.95 & 1.01 & -1.00 &  &  \\ 
&  SD (Cox) & 0.08 & 0.19 & 0.11 & 0.12 & 0.12 & 0.12 &  &  \\ 
  Likelihood & Est (joint) & -0.50 & 0.69 & 1.00 & -1.00 & 1.01 & -1.01 & 1.00 & 1.01 \\ 
&  SD (joint) & 0.03 & 0.04 & 0.06 & 0.06 & 0.05 & 0.05 & 0.03 & 0.05 \\ 
&  Est (IP-CC) & -0.50 & 0.70 & 1.00 & -1.00 & 1.01 & -1.01 & 1.00 & 1.00 \\ 
&  SD (IP-CC) & 0.03 & 0.07 & 0.06 & 0.07 & 0.07 & 0.07 & 0.06 & 0.09 \\ 
\hline
\end{tabular}
\end{table}

\begin{table}
\centering
\caption{Estimates (Est) and empirical standard deviations (SDs) based on 500 replications for $n_0=500, n_1=1000, n_2 = 500$. In the population, $(X_1, X_2)^T$ are  multivariate normally distributed with mean $(0,0)$, $Var(X_1)=Var(X_2)=1$ and  $Cov(X_1,X_2)=0.5$.  
}
\begin{tabular}{ll rr rr rr rr}
 \hline
\multicolumn{1}{c}{Method}& & \multicolumn{1}{c}{$\alpha$} &\multicolumn{1}{c}{$\nu$} & \multicolumn{1}{c}{$\beta_1$= 1} & \multicolumn{1}{c}{$\beta_2$= -1} & \multicolumn{1}{c}{$\gamma_1$= 1} & \multicolumn{1}{c}{$\gamma_2$= -1} & \multicolumn{1}{c}{$k_1$= 1} & \multicolumn{1}{c}{$k_2$= 1} \\ 
		\cmidrule{1-10}
	& & \multicolumn{8}{c}{10\% censoring}\\
\cmidrule{2-10}
 Two-step & Est (EM) & 0.18 & -0.08 & 1.01 & -1.01 & 1.02 & -1.02 &  &  \\ 
&  SD (EM) & 0.03 & 0.03 & 0.06 & 0.07 & 0.04 & 0.04 &  &  \\ 
&Est (Cox) & 0.19 & 0.00 & 1.00 & -1.00 & 1.00 & -1.00 &  &  \\ 
&  SD (Cox) & 0.03 & 0.04 & 0.06 & 0.06 & 0.04 & 0.04 &  &  \\ 
  Likelihood &  Est (joint) & 0.19 & 0.00 & 1.00 & -1.00 & 1.00 & -1.00 & 1.00 & 1.00 \\ 
&  SD (joint) & 0.03 & 0.03 & 0.06 & 0.06 & 0.03 & 0.03 & 0.02 & 0.03 \\ 
&  Est (IP-CC) & 0.19 & -0.00 & 1.00 & -1.00 & 1.02 & -1.02 & 1.01 & 1.01 \\ 
&  SD (IP-CC) & 0.03 & 0.10 & 0.06 & 0.07 & 0.10 & 0.10 & 0.09 & 0.13 \\ 
\cmidrule{1-10}
		& & \multicolumn{8}{c}{50\% censoring}\\
\cmidrule{2-10}
Two-step  & Est (EM) & 0.16 & -0.23 & 1.04 & -1.04 & 1.00 & -1.00 &  &  \\ 
&  SD (EM) & 0.03 & 0.09 & 0.06 & 0.07 & 0.05 & 0.05 &  &  \\ 
&Est (Cox) & 0.19 & -0.02 & 0.99 & -1.00 & 1.00 & -1.00 &  &  \\ 
&  SD (Cox) & 0.04 & 0.07 & 0.07 & 0.07 & 0.05 & 0.05 &  &  \\ 
  Likelihood &  Est (joint) & 0.19 & 0.00 & 1.00 & -1.00 & 1.01 & -1.00 & 1.00 & 1.00 \\ 
&  SD (joint) & 0.03 & 0.03 & 0.06 & 0.07 & 0.04 & 0.04 & 0.02 & 0.04 \\ 
&  Est (IP-CC) & 0.19 & 0.00 & 1.00 & -1.00 & 1.02 & -1.02 & 1.01 & 1.01 \\ 
&  SD (IP-CC) & 0.03 & 0.10 & 0.06 & 0.07 & 0.10 & 0.11 & 0.09 & 0.13 \\ 

\cmidrule{1-10}
 	   &  & \multicolumn{8}{c}{90\% censoring}\\ 
\cmidrule{2-10}
 Two-step& Est (EM) & 0.24 & -0.56 & 0.93 & -0.93 & 0.87 & -0.87 &  &  \\ 
&  SD (EM) & 0.07 & 0.23 & 0.10 & 0.10 & 0.14 & 0.14 &  & \\ 
&Est (Cox) & 0.24 & -0.12 & 0.92 & -0.92 & 1.01 & -1.01 &  & \\ 
&  SD (Cox) & 0.06 & 0.26 & 0.09 & 0.09 & 0.12 & 0.11 &  &  \\ 
  Likelihood & Est (joint) & 0.19 & -0.00 & 1.00 & -1.00 & 1.01 & -1.01 & 1.00 & 1.01 \\ 
&  SD (joint) & 0.03 & 0.04 & 0.06 & 0.07 & 0.06 & 0.06 & 0.03 & 0.05 \\ 
&  Est (IP-CC) & 0.19 & 0.00 & 1.00 & -1.00 & 1.02 & -1.02 & 1.01 & 1.01 \\ 
&  SD (IP-CC) & 0.03 & 0.10 & 0.06 & 0.07 & 0.10 & 0.11 & 0.09 & 0.14 \\ 
\hline
\end{tabular}
\end{table}

\begin{table}
\centering
\caption{Estimates (Est) and empirical standard deviations (SDs) based on 500 replications for $n_0=500, n_1=500, n_2 = 500$. In the population, $(X_1, X_2)^T$ are  multivariate normally distributed with mean $(0,0)$, $Var(X_1)=Var(X_2)=1$ and  $Cov(X_1,X_2)=0.5$.  
}
\begin{tabular}{ll rr rr rr rr}
 \hline
\multicolumn{1}{c}{Method}& & \multicolumn{1}{c}{$\alpha$} &\multicolumn{1}{c}{$\nu$} & \multicolumn{1}{c}{$\beta_1$= 0} & \multicolumn{1}{c}{$\beta_2$= 0} & \multicolumn{1}{c}{$\gamma_1$= 1} & \multicolumn{1}{c}{$\gamma_2$= -1} & \multicolumn{1}{c}{$k_1$= 1} & \multicolumn{1}{c}{$k_2$= 1} \\ 
		\cmidrule{1-10}
	& & \multicolumn{8}{c}{10\% censoring}\\
\cmidrule{2-10}
 Two-step & Est (EM) & 0.00 & -0.66 & -0.02 & 0.02 & 0.96 & -0.96 &  &  \\ 
&  SD (EM) & 0.00 & 0.16 & 0.07 & 0.07 & 0.08 & 0.08 &  &  \\ 
&Est (Cox) & 0.00 & -0.49 & -0.00 & -0.00 & 1.00 & -1.00 &  &  \\ 
&  SD (Cox) & 0.00 & 0.05 & 0.07 & 0.07 & 0.05 & 0.05 &  &  \\ 
  Likelihood &  Est (joint) & 0.00 & -0.48 & -0.01 & 0.00 & 0.99 & -0.99 & 1.01 & 1.00 \\ 
&  SD (joint) &  0.00 & 0.04 & 0.06 & 0.07 & 0.04 & 0.04 & 0.03 & 0.04 \\ 
&  Est (IP-CC) & 0.00 & -0.49 & -0.01 & 0.01 & 1.00 & -1.00 & 1.04 & 1.05 \\ 
&  SD (IP-CC) & 0.00 & 0.11 & 0.07 & 0.07 & 0.10 & 0.10 & 0.09 & 0.14 \\ 
\cmidrule{1-10}
		& & \multicolumn{8}{c}{50\% censoring}\\
\cmidrule{2-10}
 Two-step & Est (EM) & 0.00 & -0.67 & -0.02 & 0.02 & 0.98 & -0.98 &  &  \\ 
&  SD (EM) & 0.00 & 0.07 & 0.07 & 0.07 & 0.06 & 0.06 &  &  \\ 
&Est (Cox) & 0.00 & -0.50 & -0.00 & -0.00 & 1.01 & -1.00 &  &  \\ 
&  SD (Cox) & 0.00 & 0.06 & 0.07 & 0.07 & 0.07 & 0.07 &  &  \\
  Likelihood &  Est (joint) & 0.00 & -0.48 & -0.00 & -0.00 & 1.01 & -1.01 & 1.00 & 1.00 \\ 
&  SD (joint) & 0.00 & 0.05 & 0.07 & 0.07 & 0.05 & 0.05 & 0.03 & 0.04 \\ 
&  Est (IP-CC) & 0.00 & -0.48 & -0.00 & -0.00 & 1.02 & -1.02 & 1.01 & 1.01 \\ 
&  SD (IP-CC) & 0.00 & 0.11 & 0.07 & 0.07 & 0.10 & 0.11 & 0.09 & 0.14 \\ 
\cmidrule{1-10}
 	   &  & \multicolumn{8}{c}{90\% censoring}\\ 
\cmidrule{2-10}
 Two-step & Est (EM) & -0.00 & -0.96 & -0.08 & 0.07 & 0.88 & -0.88 &  &  \\ 
&  SD (EM) & 0.00 & 0.14 & 0.07 & 0.07 & 0.08 & 0.08 &  &  \\ 
&Est (Cox) & -0.00 & -0.63 & -0.03 & 0.02 & 1.01 & -1.02 &  &  \\ 
&  SD (Cox) & 0.00 & 0.20 & 0.08 & 0.08 & 0.14 & 0.14 &  &  \\ 
  Likelihood &   Est (joint) & 0.00 & -0.49 & -0.00 & -0.00 & 1.01 & -1.02 & 1.01 & 1.01 \\ 
&  SD (joint) & 0.00 & 0.06 & 0.07 & 0.07 & 0.06 & 0.07 & 0.04 & 0.06 \\ 
&  Est (IP-CC) & 0.00 & -0.48 & -0.00 & -0.00 & 1.02 & -1.02 & 1.01 & 1.00 \\ 
&  SD (IP-CC) & 0.00 & 0.11 & 0.07 & 0.07 & 0.10 & 0.11 & 0.10 & 0.14 \\ 
\hline
\end{tabular}
\end{table}

\begin{table}
\centering
\caption{Estimates (Est) and empirical standard deviations (SDs) based on 500 replications for $n_0=500, n_1=500, n_2 = 1000$. In the population, $(X_1, X_2)^T$ are  multivariate normally distributed with mean $(0,0)$, $Var(X_1)=Var(X_2)=1$ and  $Cov(X_1,X_2)=0.5$.  
}
\begin{tabular}{ll rr rr rr rr}
 \hline
\multicolumn{1}{c}{Method}& & \multicolumn{1}{c}{$\alpha$} &\multicolumn{1}{c}{$\nu$} & \multicolumn{1}{c}{$\beta_1$= 0} & \multicolumn{1}{c}{$\beta_2$= 0} & \multicolumn{1}{c}{$\gamma_1$= 1} & \multicolumn{1}{c}{$\gamma_2$= -1} & \multicolumn{1}{c}{$k_1$= 1} & \multicolumn{1}{c}{$k_2$= 1} \\ 
		\cmidrule{1-10}
	& & \multicolumn{8}{c}{10\% censoring}\\
\cmidrule{2-10}
 Two-step & Est (EM) & 0.00 & -0.04 & -0.05 & 0.05 & 0.93 & -0.93 &  &  \\ 
&  SD (EM) & 0.00 & 0.21 & 0.07 & 0.07 & 0.09 & 0.09 &  &  \\ 
&Est (Cox) & 0.00 & 0.21 & 0.00 & -0.00 & 1.00 & -1.00 &  &  \\ 
&  SD (Cox) & 0.00 & 0.04 & 0.06 & 0.07 & 0.05 & 0.04 & &  \\ 
  Likelihood &  Est (joint) & 0.00 & 0.21 & -0.01 & 0.00 & 0.99 & -0.99 & 1.00 & 1.01 \\ 
&  SD (joint) & 0.00 & 0.04 & 0.06 & 0.07 & 0.03 & 0.04 & 0.02 & 0.04 \\ 
&  Est (IP-CC) & 0.00 & 0.19 & -0.02 & 0.02 & 0.99 & -0.99 & 1.03 & 1.05 \\ 
&  SD (IP-CC) & 0.00 & 0.08 & 0.06 & 0.07 & 0.07 & 0.07 & 0.07 & 0.10 \\ 
\cmidrule{1-10}
		& & \multicolumn{8}{c}{50\% censoring}\\
\cmidrule{2-10}
 Two-step & Est (EM) & -0.00 & -0.03 & -0.04 & 0.04 & 0.97 & -0.97 &  &  \\ 
&  SD (EM) & 0.00 & 0.08 & 0.06 & 0.07 & 0.04 & 0.05 &  &  \\ 
&Est (Cox) & 0.00 & 0.21 & -0.00 & -0.00 & 1.00 & -1.00 &  &  \\ 
&  SD (Cox) & 0.00 & 0.05 & 0.06 & 0.07 & 0.06 & 0.05 &  &  \\
  Likelihood &  Est (joint) & 0.00 & 0.21 & -0.00 & 0.00 & 1.00 & -1.00 & 1.00 & 1.00 \\ 
&  SD (joint) & 0.00 & 0.04 & 0.06 & 0.07 & 0.04 & 0.04 & 0.03 & 0.04 \\ 
&  Est (IP-CC) & 0.00 & 0.21 & -0.01 & 0.00 & 1.01 & -1.01 & 1.01 & 1.01 \\ 
&  SD (IP-CC) & 0.00 & 0.08 & 0.06 & 0.07 & 0.07 & 0.07 & 0.07 & 0.10 \\ 
\cmidrule{1-10}
 	   &  & \multicolumn{8}{c}{90\% censoring}\\ 
\cmidrule{2-10}
 Two-step & Est (EM) & -0.00 & -0.08 & -0.06 & 0.05 & 0.95 & -0.95 &  &  \\ 
&  SD (EM) & 0.00 & 0.07 & 0.06 & 0.07 & 0.08 & 0.08 &  &  \\ 
&Est (Cox) & -0.00 & 0.13 & -0.00 & 0.00 & 1.01 & -1.01 &  &  \\ 
&  SD (Cox) & 0.00 & 0.12 & 0.08 & 0.08 & 0.12 & 0.11 &  &  \\ 
  Likelihood &  Est (joint) & 0.00 & 0.21 & -0.00 & -0.00 & 1.01 & -1.01 & 1.00 & 1.00 \\ 
&  SD (joint) & 0.00 & 0.05 & 0.06 & 0.07 & 0.05 & 0.05 & 0.03 & 0.05 \\ 
&  Est (IP-CC) & 0.00 & 0.21 & -0.00 & -0.00 & 1.01 & -1.01 & 1.01 & 1.00 \\ 
&  SD (IP-CC) & 0.00 & 0.08 & 0.06 & 0.07 & 0.07 & 0.07 & 0.07 & 0.10 \\ 
\hline
\end{tabular}
\end{table}

\begin{table}
\centering
\caption{Estimates (Est) and empirical standard deviations (SDs) based on 500 replications for $n_0=500, n_1=1000, n_2 = 500$. In the population, $(X_1, X_2)^T$ are  multivariate normally distributed with mean $(0,0)$, $Var(X_1)=Var(X_2)=1$ and  $Cov(X_1,X_2)=0.5$.  
}
\begin{tabular}{ll rr rr rr rr}
 \hline
\multicolumn{1}{c}{Method}& & \multicolumn{1}{c}{$\alpha$} &\multicolumn{1}{c}{$\nu$} & \multicolumn{1}{c}{$\beta_1$= 0} & \multicolumn{1}{c}{$\beta_2$= 0} & \multicolumn{1}{c}{$\gamma_1$= 1} & \multicolumn{1}{c}{$\gamma_2$= -1} & \multicolumn{1}{c}{$k_1$= 1} & \multicolumn{1}{c}{$k_2$= 1} \\ 
		\cmidrule{1-10}
	& & \multicolumn{8}{c}{10\% censoring}\\
\cmidrule{2-10}
 Two-step & Est (EM) & 0.69 & -0.60 & -0.01 & 0.01 & 0.97 & -0.97 &  &  \\ 
&  SD (EM) & 0.00 & 0.13 & 0.06 & 0.06 & 0.06 & 0.06 &  &  \\ 
&Est (Cox) & 0.69 & -0.49 & 0.00 & -0.00 & 1.00 & -1.00 &  &  \\ 
 & SD (Cox) & 0.00 & 0.04 & 0.06 & 0.06 & 0.04 & 0.04 &  &  \\ 
  Likelihood &  Est (joint) & 0.69 & -0.48 & -0.00 & -0.00 & 0.99 & -0.99 & 1.00 & 1.00 \\ 
&  SD (joint) & 0.00 & 0.04 & 0.06 & 0.06 & 0.03 & 0.03 & 0.02 & 0.03 \\ 
&  Est (IP-CC) & 0.69 & -0.49 & -0.01 & 0.01 & 1.01 & -1.00 & 1.04 & 1.05 \\ 
&  SD (IP-CC) & 0.00 & 0.10 & 0.06 & 0.06 & 0.10 & 0.10 & 0.10 & 0.15 \\ 
\cmidrule{1-10}
		& & \multicolumn{8}{c}{50\% censoring}\\
\cmidrule{2-10}
 Two-step & Est (EM) & 0.69 & -0.60 & -0.00 & 0.00 & 0.99 & -0.99 &  &  \\ 
&  SD (EM) & 0.00 & 0.05 & 0.06 & 0.06 & 0.05 & 0.05 &  &  \\ 
&Est (Cox) & 0.69 & -0.50 & -0.00 & -0.00 & 1.00 & -1.00 &  &  \\ 
&  SD (Cox) & 0.00 & 0.06 & 0.06 & 0.06 & 0.05 & 0.05 &  &  \\   
  Likelihood &  Est (joint) & 0.69 & -0.48 & -0.00 & -0.00 & 1.00 & -1.00 & 1.00 & 1.00 \\ 
&  SD (joint) & 0.00 & 0.04 & 0.06 & 0.06 & 0.04 & 0.04 & 0.02 & 0.03 \\ 
&  Est (IP-CC) & 0.69 & -0.48 & -0.00 & -0.00 & 1.02 & -1.02 & 1.02 & 1.01 \\ 
&  SD (IP-CC) & 0.00 & 0.11 & 0.06 & 0.06 & 0.11 & 0.11 & 0.10 & 0.15 \\ 
\cmidrule{1-10}
 	   &  & \multicolumn{8}{c}{90\% censoring}\\ 
\cmidrule{2-10}
Two-step & Est (EM) & 0.69 & -0.83 & -0.03 & 0.03 & 0.92 & -0.92 &  &  \\ 
&  SD (EM) & 0.00 & 0.12 & 0.06 & 0.06 & 0.08 & 0.07 &  &  \\ 
&Est (Cox) & 0.69 & -0.63 & -0.02 & 0.01 & 1.00 & -1.00 &  &  \\ 
&  SD (Cox) & 0.00 & 0.20 & 0.06 & 0.07 & 0.11 & 0.10 &  &  \\ 
 Likelihood &   Est (joint) & 0.69 & -0.48 & -0.00 & -0.00 & 1.01 & -1.01 & 1.00 & 1.00 \\ 
&  SD (joint) & 0.00 & 0.05 & 0.06 & 0.06 & 0.06 & 0.06 & 0.03 & 0.05 \\ 
&  Est (IP-CC) & 0.69 & -0.48 & -0.00 & -0.00 & 1.02 & -1.02 & 1.01 & 1.01 \\ 
&  SD (IP-CC) & 0.00 & 0.11 & 0.06 & 0.06 & 0.11 & 0.11 & 0.10 & 0.15 \\ 
\hline
\end{tabular}
\end{table}

\begin{table}
\centering
\caption{Estimates (Est) and empirical standard deviations (SDs) based on 500 replications for $n_0=500, n_1=500, n_2 = 500$ and step function baseline with $ \lambda_0(t)= 10^{-4}$ in $[0,7]$, $10^{-5}$ in $(7,14]$, $2\times 10^{-4}$ in $(14, 21]$, $0.5\times 10^{-4}$ in $(21, 30]$. In the population, $(X_1, X_2)^T$ are  multivariate normally distributed with mean $(0,0)$, $Var(X_1)=Var(X_2)=1$ and  $Cov(X_1,X_2)=0.5$.  
}
\begin{tabular}{ll rr rr rr rr}
 \hline
\multicolumn{1}{c}{Method}& & \multicolumn{1}{c}{$\alpha$} &\multicolumn{1}{c}{$\nu$} & \multicolumn{1}{c}{$\beta_1$= 1} & \multicolumn{1}{c}{$\beta_2$= -1} & \multicolumn{1}{c}{$\gamma_1$= 1} & \multicolumn{1}{c}{$\gamma_2$= -1} & \multicolumn{1}{c}{$k_1$= 1} & \multicolumn{1}{c}{$k_2$= 1} \\ 
		\cmidrule{1-10}
	& & \multicolumn{8}{c}{10\% censoring}\\
\cmidrule{2-10}
Two-step & Est (EM) & -0.52 & -2.13 & 1.04 & -1.04 & 0.80 & -0.80 &  &  \\ 
&  SD (EM) & 0.04 & 0.04 & 0.07 & 0.07 & 0.05 & 0.05 &  &  \\ 
&Est (Cox) & -0.52 & -2.12 & 1.04 & -1.04 & 0.79 & -0.80 &  &  \\ 
&  SD (Cox) & 0.04 & 0.04 & 0.07 & 0.07 & 0.05 & 0.05 &  &  \\ 
Likelihood & Est (joint) & -0.51 & -2.09 & 1.03 & -1.04 & 0.74 & -0.74 & 0.97 & 8.24 \\ 
&  SD (joint) & 0.04 & 0.04 & 0.07 & 0.07 & 0.04 & 0.04 & 0.02 & 0.36 \\ 
&  Est (IP-CC) & -0.51 & -2.17 & 1.04 & -1.04 & 0.78 & -0.78 & 0.98 & 9.21 \\ 
&  SD (IP-CC) & 0.04 & 0.11 & 0.08 & 0.08 & 0.09 & 0.09 & 0.12 & 1.41 \\ 
\cmidrule{1-10}
		& & \multicolumn{8}{c}{50\% censoring}\\
\cmidrule{2-10}
Two-step & Est (EM) & -0.53 & -2.16 & 1.05 & -1.06 & 0.85 & -0.85 &  &  \\ 
&  SD (EM) & 0.04 & 0.05 & 0.07 & 0.07 & 0.06 & 0.06 &  &  \\ 
&Est (Cox) & -0.53 & -2.16 & 1.05 & -1.06 & 0.87 & -0.88 &  &  \\ 
 & SD (Cox) & 0.04 & 0.05 & 0.07 & 0.08 & 0.07 & 0.07 & & \\ 
Likelihood &  Est (joint) & -0.53 & -2.14 & 1.06 & -1.06 & 0.81 & -0.81 & 0.94 & 8.97 \\ 
&  SD (joint) & 0.04 & 0.04 & 0.07 & 0.07 & 0.05 & 0.05 & 0.03 & 0.53 \\ 
&  Est (IP-CC) & -0.51 & -2.16 & 1.04 & -1.04 & 0.79 & -0.79 & 0.97 & 9.23 \\ 
&  SD (IP-CC) & 0.04 & 0.12 & 0.08 & 0.08 & 0.10 & 0.09 & 0.12 & 1.44 \\ 
\cmidrule{1-10}
 	   &  & \multicolumn{8}{c}{90\% censoring}\\ 
\cmidrule{2-10}
Two-step & Est (EM) & -0.50 & -2.22 & 1.01 & -1.02 & 0.78 & -0.78 &  &  \\ 
&  SD (EM) & 0.06 & 0.18 & 0.08 & 0.09 & 0.17 & 0.17 &  &  \\ 
&Est (Cox) & -0.51 & -2.24 & 1.02 & -1.03 & 0.93 & -0.94 &  &  \\ 
&  SD (Cox) & 0.05 & 0.14 & 0.08 & 0.09 & 0.15 & 0.15 &  &  \\ 
Likelihood &  Est (joint) & -0.53 & -2.16 & 1.05 & -1.06 & 0.83 & -0.83 & 0.94 & 9.57 \\ 
&  SD (joint) & 0.04 & 0.06 & 0.07 & 0.08 & 0.07 & 0.07 & 0.04 & 0.87 \\ 
&  Est (IP-CC) & -0.52 & -2.15 & 1.04 & -1.05 & 0.80 & -0.80 & 0.93 & 9.34 \\ 
&  SD (IP-CC) & 0.04 & 0.12 & 0.07 & 0.08 & 0.10 & 0.10 & 0.12 & 1.54 \\ 
\hline
\end{tabular}
\end{table}

\begin{table}
\centering
\caption{Estimates (Est) and empirical standard deviations (SDs) based on 500 replications for $n_0=500, n_1=500, n_2 = 500$ and step function baseline with $ \lambda_0(t)= 10^{-5}$ in $[0,7]$, $2.0\times 10^{-4}$ in $(7,14]$, $10^{-5}$ in $(14, 21]$, $2.0\times 10^{-4}$ in $(21, 30]$. In the population, $(X_1, X_2)^T$ are  multivariate normally distributed with mean $(0,0)$, $Var(X_1)=Var(X_2)=1$ and  $Cov(X_1,X_2)=0.5$.  
}
\begin{tabular}{ll rr rr rr rr}
 \hline
\multicolumn{1}{c}{Method}& & \multicolumn{1}{c}{$\alpha$} &\multicolumn{1}{c}{$\nu$} & \multicolumn{1}{c}{$\beta_1$= 1} & \multicolumn{1}{c}{$\beta_2$= -1} & \multicolumn{1}{c}{$\gamma_1$= 1} & \multicolumn{1}{c}{$\gamma_2$= -1} & \multicolumn{1}{c}{$k_1$= 1} & \multicolumn{1}{c}{$k_2$= 1} \\ 
		\cmidrule{1-10}
	& & \multicolumn{8}{c}{10\% censoring}\\
\cmidrule{2-10}
Two-step & Est (EM) & -0.52 & -2.89 & 1.03 & -1.04 & 0.81 & -0.81 &  &  \\ 
&  SD (EM) & 0.04 & 0.04 & 0.07 & 0.08 & 0.05 & 0.05 &  &  \\ 
&Est (Cox) & -0.52 & -2.90 & 1.03 & -1.03 & 0.81 & -0.81 &  & \\ 
&  SD (Cox) & 0.04 & 0.03 & 0.07 & 0.08 & 0.05 & 0.05 &  &  \\
Likelihood &  Est (joint) & -0.51 & -2.87 & 1.02 & -1.02 & 0.68 & -0.68 & 2.07 & 15.92 \\ 
&  SD (joint) & 0.04 & 0.03 & 0.07 & 0.07 & 0.04 & 0.04 & 0.07 & 0.35 \\ 
&  Est (IP-CC) & -0.52 & -2.89 & 1.03 & -1.03 & 0.74 & -0.74 & 2.08 & 16.42 \\ 
&  SD (IP-CC) & 0.04 & 0.06 & 0.07 & 0.08 & 0.11 & 0.11 & 0.23 & 1.07 \\ 
\cmidrule{1-10}
		& & \multicolumn{8}{c}{50\% censoring}\\
\cmidrule{2-10}
Two-step & Est (EM) & -0.53 & -2.90 & 1.04 & -1.04 & 0.86 & -0.86 &  &  \\ 
&  SD (EM) & 0.04 & 0.06 & 0.07 & 0.08 & 0.06 & 0.06 &  &  \\ 
&Est (Cox) & -0.52 & -2.91 & 1.04 & -1.04 & 0.87 & -0.87 &  &  \\ 
 & SD (Cox) & 0.04 & 0.04 & 0.07 & 0.08 & 0.07 & 0.06 &  &  \\ 
Likelihood &  Est (joint) & -0.51 & -2.88 & 1.02 & -1.02 & 0.68 & -0.68 & 2.05 & 16.10 \\ 
&  SD (joint) & 0.04 & 0.03 & 0.07 & 0.07 & 0.05 & 0.05 & 0.10 & 0.44 \\ 
&  Est (IP-CC) & -0.52 & -2.89 & 1.03 & -1.03 & 0.74 & -0.74 & 2.07 & 16.45 \\ 
&  SD (IP-CC) & 0.04 & 0.06 & 0.07 & 0.08 & 0.11 & 0.11 & 0.23 & 1.08 \\ 
\cmidrule{1-10}
 	   &  & \multicolumn{8}{c}{90\% censoring}\\ 
\cmidrule{2-10}
Two-step & Est (EM) & -0.51 & -2.95 & 1.02 & -1.03 & 0.79 & -0.79 &  &  \\ 
&  SD (EM) & 0.05 & 0.11 & 0.08 & 0.08 & 0.16 & 0.16 &  &  \\ 
&Est (Cox) & -0.52 & -2.96 & 1.03 & -1.04 & 0.94 & -0.94 &  &  \\ 
&  SD (Cox) & 0.04 & 0.08 & 0.07 & 0.08 & 0.14 & 0.14 &  &  \\ 
Likelihood &  Est (joint) & -0.53 & -2.84 & 1.04 & -1.04 & 0.67 & -0.67 & 1.72 & 16.00 \\ 
&  SD (joint) & 0.04 & 0.04 & 0.07 & 0.08 & 0.07 & 0.07 & 0.09 & 0.83 \\ 
&  Est (IP-CC) & -0.52 & -2.89 & 1.03 & -1.03 & 0.75 & -0.75 & 2.06 & 16.55 \\ 
&  SD (IP-CC) & 0.04 & 0.07 & 0.07 & 0.08 & 0.11 & 0.11 & 0.23 & 1.11 \\ 
\hline
\end{tabular}
\end{table}

\end{document}